\documentstyle[aaspp4]{article}        

\def\der#1#2{{\partial#1\over\partial#2}}
\def\ders#1#2#3{{\partial^2#1\over\partial#2\partial#3}}
\def\derss#1#2{{\partial^2#1\over\partial#2^2}}
\newcommand{\x}{{{\bf x}}}
\newcommand{\etal}{{\it et al.\/} }
\newcommand{\be}{\begin{equation} }
\newcommand{\ee}{\end{equation} }
\newcommand{\bea}{\begin{eqnarray} }
\newcommand{\eea}{\end{eqnarray} }
\newcommand{\un}{{\left (1 \right )}  }
\newcommand{\du}{{\left (2 \right )}  }
\newcommand{\ud}{{\left (12 \right )}  }
\newcommand{\uu}{{\left (11 \right )}  }
\newcommand{\cd}{{{\cal D}}}
\lefthead{Shapiro \& Zane}
\righthead{BAR MODE INSTABILITY IN RELATIVISTIC ROTATING STARS}

\begin{document}

\title{BAR MODE INSTABILITY IN RELATIVISTIC ROTATING STARS:\\
A POST--NEWTONIAN TREATMENT}
\author{Stuart L. Shapiro\altaffilmark{1} and Silvia Zane\altaffilmark{2}}
\affil{Department of Physics, Loomis Laboratory for 
Physics, University of Illinois at Urbana--Champaign, Urbana, Illinois, 61801}

\altaffiltext{1}{Department of Astronomy and NCSA, University of Illinois 
at Urbana--Champaign, Urbana, Illinois, 61801} 

\altaffiltext{2}{International School for Advanced Studies SISSA/ISAS, 
Trieste, Italy; \\ \null \hskip 0.5 truecm Nuclear and Astrophysics 
Laboratory, University of Oxford, Oxford, England}
 
\begin{abstract}

We construct analytic models of incompressible, uniformly rotating stars 
in post--Newtonian (PN) gravity and evaluate their stability against 
nonaxisymmetric bar modes. We model the PN configurations by homogeneous 
triaxial ellipsoids and employ an energy variational principle to 
determine their equilibrium shape and stability. The spacetime metric is 
obtained by solving Einstein's equations of general relativity in 3+1 ADM 
form. We 
use an approximate subset of these equations well--suited to numerical 
integration in the case of strong field, three dimensional configurations 
in quasi--equilibrium. 
However, the adopted equations are exact at PN order, where they admit an 
analytic solution for homogeneous ellipsoids. We obtain this solution for 
the metric, as well as analytic functionals for the conserved global 
quantities, $M$, $M_0$ and $J$. 

We present sequences of axisymmetric, rotating equilibria of constant 
density and rest mass parametrized by their eccentricity. These 
configurations represent the PN generalization 
of Newtonian Maclaurin spheroids, which we compare to other PN and full 
relativistic incompressible equilibrium sequences constructed by previous 
investigators. We employ the variational principle to 
consider nonaxisymmetric 
ellipsoidal deformations of the configurations, holding the angular 
momentum constant and the rotation uniform. We locate the point along 
each sequence at which these Jacobi--like bar modes will be driven 
secularly unstable by 
the presence of a dissipative agent like viscosity. We find that the 
value of the eccentricity, as well as related ratios like $\Omega^2/(\pi 
\rho_0)$ and $T/|W|$ (= rotational 
kinetic energy / gravitational potential energy), defined invariantly, 
all increase at the onset of instability 
as the stars become more relativistic. Since higher degrees of rotation 
are required to trigger a 
viscosity--driven bar mode instability as the stars 
become more compact, the effect of general relativity is to 
weaken the instability, at least to PN order. This behavior is in stark 
contrast to that found recently for secular instability via 
nonaxisymmetric, Dedekind--like modes driven by gravitational radiation. 
These findings support the suggestion that in general relativity 
nonaxisymmetric 
modes driven unstable by viscosity no longer coincide with those driven 
unstable by gravitational radiation.

\end{abstract}

\keywords{gravitation --- relativity --- instabilities --- 
stars: neutron --- stars: rotation}

\section{Introduction}

The identification of nonaxisymmetric modes of 
instability in rapidly rotating equilibrium configurations is a classic 
problem. Although a considerable amount of work has been done in 
Newtonian theory, where a number of results are well established 
(see Chandrasekhar 1969a, hereafter Ch69 and section 2 for a summary 
and references), only a 
few investigations have been carried out so far in the context of general 
relativity. 
The first three dimensional (3D) perturbation computations 
in full general relativity to identify nonaxisymmetric instabilities driven by 
gravitational 
radiation have been carried out recently 
by Stergioulas \& Friedman (1997, hereafter SF) and Stergioulas (1997). 
A earlier numerical investigation of the effects of relativity on the 
viscosity--driven bar mode instability was presented by 
Bonazzola, Frieben \& Gourgoulhon (1996, hereafter BFG) and these results 
have been recently corroborated by a more detailed analysis 
(Bonazzola, Frieben \& Gourgoulhon 1997). 
SF solved the coupled set of perturbed field 
equations, modeling uniformly rotating stars by a polytropic equation of 
state (EOS) and 
adopting the Friedman \& Schutz (1995) criterion for the onset of the 
nonaxisymmetric instability to gravitational radiation dissipation. 
According to this criterion, a nonaxisymmetric mode becomes unstable when 
its frequency, as measured by an observer at infinity, vanishes. 
They find that relativistic models are unstable to nonaxisymmetric 
modes for significantly smaller degrees of rotation than for 
corresponding Newtonian models. The destabilizing effect of relativity is 
most striking in the case of the 
$m=2$ mode, which can become unstable even for soft polytropes of index 
$\gamma \geq \gamma_{crit} = 1.77$ ($n \leq 1.3$), while the critical 
index in Newtonian theory is $\gamma_{crit} = 2.238$ (Jeans 1919, 1928; 
James 1964). 
This behaviour is in agreement with results of a 
semianalytical, PN analysis previously presented by 
Cutler (1991) and Cutler \& Lindblom (1992, see also Lindblom 1995, Yoshida 
\& Eriguchi 1997). In particular, Cutler (1991) derived the PN 
corrections to the 
pulsational modes of uniformly rotating stars. The resulting expressions 
were then used by Cutler \& Lindblom (1992) to evaluate the critical 
value of the 
star angular velocity, $\Omega_{crit}$, where the frequency of the mode 
passes through zero. This is the critical value at which, in absence of 
viscosity, these modes are unstable to the emission of gravitational 
radiation. By considering the $\gamma = 2 $ polytrope, 
they found that PN effects lowers by up to 10 \% the values of 
$\Omega_{crit}$, concluding that PN effects tend to make the 
gravitational--radiation instability more important. 
However, as noted by SF, the results of BFG, 
who investigated the effects of general relativity 
on the $m=2$, viscosity--driven ``bar'' mode instability, seem to 
suggest the opposite effect. BFG's method consists of 
perturbing a stationary, axisymmetric configuration, 
obtained from a two dimensional numerical simulation, and retaining 
only the dominant terms in the 
nonaxisymmetric relativistic perturbation equations. 
BFG found that the critical adiabatic index for the bar instability 
becomes higher than the value of James as the configuration becomes 
increasingly relativistic. 
This behaviour suggests that relativistic 
effects tend to stabilize the configurations. Accordingly, SF concluded 
that, in general relativity, the point of onset of the viscosity--driven and 
the gravitational radiation--driven $m=2$ modes may no longer coincide as 
they do in Newtonian theory, and 
that the effect of relativity seems to be very different in the two cases.
The main improvement presented in Bonazzola, Frieben \& Gourgoulhon 
(1997) over their previous study consists in the fully 
three dimensional treatment of the shift vector. With this more detailed 
analysis, the stabilization of relativistic configurations is not only 
confirmed, but also strongly enhanced. 
Both SF's and BFG's findings are based on a numerical solution 
of perturbation equations and represent the only attempts to solve the 
relativistic problem in 3D to date. 

In this paper we reconsider the problem of relativistic rotating 
equilibria and bar mode instabilities from an analytic point of view. 
Specifically, we extend the earlier Newtonian treatment of 
Lai, Rasio \& Shapiro (1993a, hereafter LRS), which is based on triaxial 
ellipsoid models of rotating stars and an energy variational principle, 
to post--Newtonian (PN) gravitation. 
We restrict our discussion to incompressible, rigidly rotating 
bodies and neglect any deviation from the ellipsoidal 
shape in the equilibrium configuration. The ellipsoidal approximation is 
exact for rotating incompressible stars in Newtonian theory, but it is 
only approximate for PN configurations. However, our formalism allows us to 
derive the analytic functionals for the main global parameters 
characterizing a rotating configuration (total mass--energy, rest mass 
and angular 
momentum). By applying a energy variational principle to these 
functionals, it is possible to construct equilibrium 
sequences of constant rest mass and to locate instability points along 
the sequence. 

Our PN analysis is carried out in the 
framework of a 3+1 ADM splitting of the metric (Arnowitt, Deser and Misner
1962). In this respect, our solution sets up and should provide 
an important test--bed calculation for future numerical studies of 3D 
relativistic rotating configurations. Numerical relativity in 3D is only 
in its infancy, and lacks a large
body of known solutions which it can attempt to reproduce. Part of the
complication is related to the fact that, in general 
relativity, 3D stellar systems are usually 
fully dynamical due to the generation of gravitational waves. As a 
result, it
is typically necessary to solve the full set of Einstein's equations 
to determine the behavior of a 3D system. There are many 
asymmetric systems, however, in which the intrinsic dynamical timescale 
is much shorter than the timescale for 
dissipation due to gravitational waves. Such is the case for any 
weak field, slow velocity system, or for a strong-field, high velocity system
that is only slightly perturbed from stationary equilibrium.
It is thus possible to discuss ``quasi--equilibrium'' 
configurations for such objects. Examples include 
rotating neutron stars (NSs) either with small compaction 
$M/R$ or small departure from axisymmetry, and binary neutron stars 
prior to reaching the innermost stable circular orbit (Baumgarte \etal 
1997b). 
Following the discussion by Wilson \& Matthews (1989, 1995) and Wilson 
(1990), Cook, Shapiro and Teukolsky 1992 (hereafter CST92)
have provided a simplified 
set of 3+1, ADM equations which are well--adapted for studying 
quasi--equilibrium relativistic systems. For 
strong-field objects, these equations yield precise solutions to the 
initial value (constraint) equations, and approximate instantaneous snapshots
of the object as it evolves, due to the emission of 
gravitational waves, provided it does so slowly. In this paper, we adopt the 
CST92 equations which are well suited to 
future numerical studies of the fully nonlinear equations for strong- 
field, quasi--equilibrium sources. 
We solve these equations in the PN approximation; at this 
order, the CST92 equations are exact. 
In this paper, we focus on 
the viscosity--driven, secular instability with respect to the 
formation of a barlike structure in a PN incompressible, rotating star. 
We find that PN solutions 
are unstable at rotation rates greater than in the Newtonian limit. 
Thus, in the PN treatment, nonaxisymmetric instabilities driven by 
viscosity set a less 
stringent limit on the maximum angular velocity than suggested by Newtonian 
theory. 

From the observational point of view, the issue of investigating the 
maximum spin velocity for a NS is manyfold. 
Rotating configurations subject to nonaxisymmetric instabilities could 
become important sources of gravitational waves emission and represent 
possible candidates for detection by
laser interferometers now under construction,
like LIGO, GEO and VIRGO (see e.g. Bonazzola \& Marck 1994; 
Thorne 1987; Lai \& Shapiro 1995; Schutz 1997 and reference therein). 
There are a number of plausible astrophysical situations 
in which this effect can set in. 
Numerical simulations of coalescing binary neutron stars show that 
merger into a single object is the probable end point of
close binary evolution (see e.g. Rasio \& Shapiro 1992, 1994; LRS; Lai, 
Rasio \& Shapiro 1993b, 1994a,b,c). 
The newly born object may be rotating quite rapidly (Baumgarte \etal 
1997b) and might be driven secularly unstable. 
Core collapse in massive, evolved stars or accretion-induced collapse of
white dwarfs, also has been suggested as 
observable sources of gravitational radiation (Lai \& Shapiro 1995 and 
references therein). 
Coalescing white dwarf binaries are though to be progenitors of type 
Ia supernovae (Iben \& Tutukov 1984; Yungelson \etal 1994), or, in certain 
cases, of isolated millisecond pulsars (Chen \& Leonard 1993). 
In all these scenarios, the newly born NS can 
undergo a secular or dynamical 
evolution, breaking its axisymmetry, provided that its spin 
velocity is larger than the critical value. Alternatively, accretion 
onto NSs in X--ray binary systems or in Thorne--Zytkow objects can spin 
up the compact star until reaching the 
critical value of $T/|W|$, at which point the Chandrasekhar-Friedman-Schutz
(CFS) instability (Chandrasekhar 1970; Friedman \& Schutz 1978) may drive 
a nonaxisymmetric mode unstable, powering radiation. This 
``forced gravitational emission'', is particularly appealing 
for maintaining steady, periodic emission, 
with the total amount of accreted angular momentum balancing 
the amount that is 
radiated away via gravitational radiation (Wagoner 1984; Schutz 1997). 

We point out that our analytical results assume 
incompressible and rigidly rotating bodies, and both assumptions have 
been introduced mainly to make 
an analytical treatment tractable, to PN order. 
However, for applications to NSs, we note that both assumptions  
are not entirely ``ad hoc''. 
Viscosity tends to drive NS to rigid rotation (Friedmann \& 
Ipser 1987). 
In addition, recent many body calculations suggest that the EOS of 
dense nuclear matter is relatively stiff (see, e.g. Wiringa, Fiks \& 
Fabronici 1988 and references therein). In view of these 
considerations, application of our results to realistic situations may
not be unreasonable, at least as a first approximation. 

In section 2, we review the physical problem and summarize the 
results of previous investigations. 
The variational principle of LRS is reviewed in section 3, 
while in section 4 we set up the mathematical 
problem in the PN approximation. We assemble the relevant set of 3+1 
equations for the gravitational field and present 
analytic solutions for both the metric coefficients and the global 
conserved quantities. These results are then used to build 
equilibrium sequences in section 5 and locate the bar mode instability 
point in section 6. Discussion and conclusions follow in section 7.

\section{Previous Investigations}

The problem of the equilibrium shape and stability of 
an incompressible, rigidly rotating configuration
admits an exact analytic solution in Newtonian gravitation (Ch69). 
In this case the rotating 
configuration takes in axisymmetry the equilibrium shape of a 
Maclaurin spheroid. However, nonaxisymmetric 
instabilities can develop in 
rapidly spinning spheroids when 
the ratio $T/|W|$ of the rotational kinetic to the 
gravitational potential energy 
becomes sufficiently large. At the critical value  
$T/|W|=0.1375$ the equilibrium sequence of 
MacLaurin configurations bifurcates into 
two other branches of triaxial equilibria, the Jacobi and the Dedekind 
ellipsoids. Since the Maclaurin 
spheroids are dynamically unstable only for $T/|W| > 0.2738$, the 
bifurcation point is dynamically stable. However, in presence 
of a suitable dissipative mechanism such as viscosity or 
gravitational radiation, this point becomes secularly unstable 
to a $l=2$ $m=2$ bar mode.  Secular 
instability may play an important role in limiting the maximum 
rotation of neutron stars. Viscosity dissipates 
the mechanical energy $E$, but preserves angular momentum $J$. 
Consequently, a Maclaurin spheroid undergoing a viscosity 
driven instability terminates its evolution as a 
Jacobi ellipsoid, with a lower value of 
$E$ but the same value of $J$ (and rest mass $M_0$). 
The Jacobi solution is rigidly rotating, so that the viscous dissipation 
stops once it is formed. 
Alternatively, the Dedekind ellipsoid represents a 
lower energy state (with respect to the Maclaurin 
solution) for a given circulation.
A rapidly spinning configuration can be unstable to the 
emission of gravitational radiation at the bifurcation point 
(the CFS instability).
Since this process 
does not conserve angular momentum but conserves  
circulation, under a CFS instability the 
growth of the bar mode leads to the deformation of a Maclaurin 
spheroid into a Dedekind ellipsoid. 
At this final state, which is stationary, the emission of 
gravitational waves stops. 
The competition between the Jacobi--like and  
the Dedekind--like modes 
is governed by the ratio of 
the strength of the viscous stress to the gravitational radiation 
reaction force, and this ratio depends crucially on 
the internal properties of the star. 
Although the determination of a realistic equation of state for 
NSs and, as a consequence, of the strength of the viscosity, is 
still an open issue, viscosity is thought to be more 
important for old NSs and gravitational radiation in the hottest, newly born 
objects (see e.g. BFG and references therein). 
The situation is more complicated when both viscosity and gravitational 
radiation act together, since they tend
to cancel each other, 
stabilizing the star (Lindblom \& Detweiler 1977; Lai \& Shapiro 1995). 

A number of efforts in Newtonian physics have been devoted to the extension 
of the Ch69's results to more realistic, compressible 
fluids, modeled by a polytropic 
equation of state. For rigidly rotating polytropes, bifurcation 
to triaxial configurations can only exist when the adiabatic 
index exceeds a critical value, $\gamma_{crit} = 2.238$ (Jeans 1919, 
1928; James 1964). This is because the EOS must be stiff enough to 
make the angular velocity at the bifurcation point lower than 
the limiting value $\Omega_k$ at which the centrifugal force 
balance the gravitational force at the equator (mass shedding limit). 
Moreover, Ipser \& Managan (1985) demonstrated that the $m=2$ 
Jacobi--like bifurcation point and the $m=2$ Dedekind--like 
point have the same location along uniformly rotating, polytropic 
sequences, as in the incompressible case (see also LRS; 
Lai \& Shapiro 1995). Typically, rotating equilibrium stellar models 
must be constructed numerically. Models have been constructed by a 
number of authors, using both polytropic (see e.g. 
Bodenheimer \& Ostriker 1973; Ipser \& Managan 1981; Hachisu \& 
Eriguchi 1982; Hachisu 1986a,b; see Tassoul 1978 for an extensive set 
of references) and more realistic equations of state for both white 
dwarfs and NSs (see e.g. Ostriker \& Tassoul 1969; Durisen, 1975; 
Hachisu 1986a; BFG and 
references therein).

LRS constructed triaxial ellipsoid models of rotating polytropic 
stars in Newtonian gravity, using an ellipsoidal energy variational method. 
This approach was 
originally introduced by Zel'dovich \& Novikov (1971, see also 
Shapiro \& Teukolsky 1983, hereafter ST) in the axisymmetric case, 
to investigate the stability of a polytropic star against gravitational 
collapse. The main advantage of the  
method comes from its simplicity: all results are analytic or 
quasi--analytic, lending themselves to straightforward 
physical interpretation. This is 
in part because the method deals 
directly with global, conserved quantities. When 
quantities like the total energy and the total angular momentum 
are determined along an 
equilibrium sequence, the evolution 
of the system can be tracked as it loses $E$ or $J$ 
by some quasi--static dissipative process. 
Generalizing the same approach to the triaxial 
case, LRS were able to construct 
equilibrium sequences for compressible analogues of most classical 
incompressible 
objects, like isolated Maclaurin, Jacobi, Dedekind and 
and Riemann ellipsoids and binary Roche, 
Darwin and Roche--Riemann ellipsoids. 

Neutron stars are relativistic objects and the analysis of their 
equilibrium and stability must be based necessarily on 
general relativistic models. 
The structure of a rotating axisymmetric star in general relativity has 
been investigated numerically 
by a number of authors (see e.g. Butterworth \& Ipser 1976, hereafter BI; 
Friedman, Ipser \& Parker 1986; CST92; 
Cook, Shapiro \& Teukolsky 1994a,b; 
Cook, Shapiro \& Teukolsky 1996, hereafter CST96 and references therein),
but the first fully relativistic 
computations of nonaxisymmetric instabilities have been presented only 
very recently (SF; Stergioulas 1997). From the analytical point of view, 
even less is known. 
An exact treatment of the radial oscillations of a gaseous 
mass in general relativity is possible (Chandrasekhar 1964), but a similar 
analysis of nonradial oscillations is not to be expected. 
In fact, apart from the difficulties 
associated with the solution of 
the Einstein equations 
without any presupposed symmetry, allowance 
must be made in the relativistic regime for the emission of gravitational 
waves. However, some insight into the nature of the general 
relativistic effects can be obtained by 
examining the problem in the PN approximation, 
i.e. at a level in which gravitational 
radiation plays no role. PN effects on the equilibrium of uniformly 
rotating, homogeneous bodies have been 
extensively investigated in a series of paper by Chandrasekhar (1965a,b, 
1967a,b,c, 
1969b; see also Chandrasekhar \& Nutku 1969, hereafter CN, for the PPN 
corrections) using the tensor virial formalism. 
Whenever possible in this paper, we make a 
direct comparison between our 
expressions and the corresponding PN results derived by Chandrasekhar. 
In particular, starting from the PN equations 
of hydrodynamics, he derived the equilibrium 
relation between the eccentricity and the angular velocity.
Although he obtained integral expressions for the global conserved 
quantities, he did not evaluate them or give explicit formulae 
for the PN corrections to the 
rest--mass, angular momentum  and binding energy. 
A different method, based on the solution of the PN Poisson 
equation in 
oblate spheroidal coordinates, was presented by Bardeen (1971). The 
formalism turns out to be simpler with respect to the Chandrasekhar one, 
but this approach does not allow an immediate generalization to the 
triaxial case. In 3D, the general eigenfunctions 
resulting from the separation (Lame functions) are indeed available, 
but many details about their general properties are not well studied. 
No analytic investigations of the location of the secular stability point 
in general relativity were provided in these earlier studies or, to our 
knowledge, elsewhere in the literature.

\section{The Energy Variational Method}  

In this section we briefly review the energy variational approach in 
Newtonian theory (LRS), 
to introduce the basic concepts that we use in our PN calculations. 

Consider a self--gravitating, isolated system. 
Each configuration (whether or not in equilibrium) can be specified by 
the total energy $E$ 
and a number of conserved global quantities, such 
as the rest mass $M_0$ and the total angular momentum $J$. 
Since $E$ can be always written in terms of the 
fluid density and velocity fields $\rho ( {\bf x} )$, $v ( {\bf x} )$    
\be
\label{01}
E = E \left [ \rho ( {\bf x} ), v ( {\bf x} ); M_0, J, \dots 
\right ] \, , 
\ee
the equilibrium state can be determined by extremizing this 
functional with respect to all variations in both $\rho ( {\bf x} )$ 
and $v ( {\bf x} )$, under the constraint that the conserved quantities 
are unchanged. Direct application of such a 
variational approach to a multidimensional system is a computationally 
challenge task. However, as 
discussed by LRS, a great 
simplification arises when we can replace the infinite number of degrees of 
freedom contained in $\rho ( {\bf x} )$ and $v ( {\bf x} )$ by a limited 
number of free parameters $\alpha_1, \alpha_2, \dots$. This we can often 
do for sufficiently simple systems under suitable simplifying 
assumptions. The total mass energy then becomes 
\be
\label{02}
E = E \left [ \alpha_1, \alpha_2, \dots ; M_0, J, \dots \right ] 
\ee
and the equilibrium configuration is determined by extremizing this 
functional according to 
\be
\label{03}
\der {E}{ \alpha_i } = 0 ~~~~~~ i = 1,2 \dots \quad \quad \quad \quad \quad 
{\rm (equilibrium)} \ee
under the constraint that $M_0$, $J$, \dots are conserved. 

The onset of the instability can be then 
determined from 
\be 
\label{001} 
{\rm det} \left ( \ders{E} {\alpha_i }{\alpha_j} \right ) _{eq} = 0 \, , 
\quad i,j = 1,2, \dots \, , \quad \quad {\rm (onset \, \, \,  of \, \, \,  
instability)} \ee
where the subscript `$eq$' indicates quantities evaluated along 
the equilibrium sequence. Clearly, whether the instability actually arises 
depends on the presence of a suitable dissipative mechanism which 
preserves the conservation laws assumed in the construction of the 
equilibrium model. 

Now consider a simple application: a homogeneous, uniformly rotating, 
Newtonian fluid system, with density 
$\rho_0$ and angular velocity $\Omega$. In the incompressible case, 
the internal energy vanishes and the total energy is given by 
\be 
\label{04}
E = T + W \, ,  
\ee 
where $T$ and $W$ are 
the rotational kinetic and the gravitational contributions, respectively.  
Assume that the surface of the configuration is ellipsoidal in shape. 
Then the geometry of the system is 
completely specified by the values $a_i$ ($i=1 \dots 3$) of the three 
semiaxes of the outer surface, where the pressure vanishes. However,
following LRS, 
it is more convenient to introduce an equivalent set of 
parameters defined by 
\be
\label{05}
\lambda_1 = \left ( { a_3 \over a_1 } \right )^{2/3} \ , 
\lambda_2 = \left ( { a_3 \over a_2 } \right )^{2/3} \, ,  
\ee 
and 
\be
\label{06}
R = (a_1 a_2 a_3 )^{1/3}\, . 
\ee 
Note that $R$ 
represents the radius of the spherical configuration with the same 
volume as the rotating one; it is not, in general, related to the 
equilibrium 
state.  
The gravitational potential and the kinetic energies can be written as 
\be 
\label{013}
W = 
- { 3 \over 5 } { G M_{0}^2 \over R} {I_{Ch} \over 2 \left ( a_1 a_2 a_3 
\right )^{2/3}} = - { 3 \over 5 } { G M_{0}^2 \over R } f \, ,  
\ee
\be 
\label{014}
T = { J^2 \over 2 I} = { J^2 \over 2 I_s } h \, , 
\ee
where $M_0 = 4 \pi \rho_0 R^3 /3$, $J= \Omega I$ and $I_s= 2 
M_0 R^2/5$ is the momentum of inertia of a sphere of the same 
volume. The momentum of inertia of the ellipsoid is 
$I= I_s /h$, and the two 
dimensionless ratios $f$ and $h$ are defined as 
\be 
\label{015}
f = { I_{Ch} \over  2 R^2 } = { 1 \over 2} 
\left ( 
{ A_1 \lambda_2 \over \lambda_1^2} 
+ { A_2 \lambda_1 \over \lambda_2^2} 
+ A_3 \lambda_1 \lambda_2 \right )  
\, , 
\ee
\be \label{016}
h = { 2 R^2 \over a_1^2 + a_2^2 } = { 2 \lambda_1^2 \lambda_2^2 \over 
\lambda_1^3 
+ \lambda_2^3}  
  \, . 
\ee
Here $I_{Ch} = \sum_i A_i a_i^2$ [called 
$I$ in Ch69, equation (3.15)], and the dimensionless coefficients $A_i$ 
can be calculated in terms of standard incomplete 
elliptic integrals involving only the axis ratios [see 
equations (3.33)--(3.35) in Ch69]. In the spherical limit $a_1 = a_2 = a_3 
= R$ and $f=h=1$. 

The total mass--energy may now be written as 
\be \label{ene}
E = - { 3 \over 10} {M_0^2 \over R^3} \left ( I_{Ch} + t h \right ) 
\, , 
\ee
where 
\be 
\label{t}
t = -{ 5 \over 3} {J^2 R^3 \over I_s M_0^2} = - {1 \over 4} {\Omega^2 
\over \pi \rho_0} { a_1^2 + a_2^2 \over h} \, . \ee
The equilibrium sequence can be constructed by varying $E$ according to 
\be \label{017} 
\der{E}{\lambda_1} = 
\der{E}{\lambda_2} = 0 \, , 
\ee
while holding $M_0$, $J$ constant. For an incompressible fluid $\rho_0 
=$ constant and there no variations with respect to  $R$. 
This gives \bea 
\label{018} 
0 &=& \der{E}{\lambda_1} = 
- { 3 \over 10} {M_0^2 \over R^3} \left (\der{I_{Ch}}{\lambda_1} + t 
\der{h}{\lambda_1} \right ) \\ \nonumber 
0 & = & \left ( 1 \leftrightarrow 2 \right ) \, . 
\eea
The derivatives of $h$ and $I_{Ch}$ with respect to $\lambda_1$ and 
$\lambda_2$ are reported in LRS [equations (A3), (A9)]. 
Exploiting these formulas and using expression (\ref{t}), 
conditions (\ref{018}) can be cast in the form  
\bea 
\label{019}
0 &=& {3 \over 2} a_1^2 A_1 - { 1 \over 2} I_{Ch} - { 1 \over 4} { \Omega^2 
\over \pi \rho_0 } \left ( 2 a_1^2 - a_2^2 \right )  \\ \nonumber 
0 & = & \left ( 1 \leftrightarrow 2 \right ) \, ,  
\eea
and these can be combined by adding each one to 1/2 times the other, yielding
\be 
\label{020} 
{\Omega^2 \over \pi \rho_0 } a_1^2 - 2 a_1^2 A_1 =  
{\Omega^2 \over \pi \rho_0 } a_2^2 - 2 a_2^2 A_2 = - 2 a_3^2 A_3 \, . 
\ee  
From Ch69, we write  
\bea 
\label{021} 
A_{ij} &=& A_{ji} = { A_i - A_j  \over a_j^2 - a_i^2 } \quad \quad \quad (i 
\neq j)  
\\ B_{ij}  & = & B_{ji} = A_j - a_i^2 A_{ij} \\ 
2 & = & 3 A_{ii} a_i^2 + A_{ij} a_i^2 + A_{ik} a_i^2  \quad (i \neq j \neq 
k) 
\eea
and add to each side of (\ref{020}) the quantity $2 
a_1^2 a_2^2 A_{12}$, obtaining 
\be
\label{022} 
a_1^2 \left ( 
{\Omega^2 \over \pi \rho_0 } - 2 B_{12} \right ) =  
a_2^2 \left ( {\Omega^2 \over \pi \rho_0 } - 2 B_{12} \right ) =  
2 \left ( 
a_1^2 a_2^2 A_{12} - a_3^2 A_3 \right ) \, . 
\ee
The latter equalities allow a solution with $a_1 \neq a_2$ if and only if 
\bea
\label{023} 
& & a_1^2 a_2^2 A_{12} =   a_3^2 A_3  \, , 
\\ \label{024} 
& & {\Omega^2 \over \pi \rho_0 } =  2 B_{12} \, , \quad \quad  \quad \quad
{\rm 
(Jacobi~ellipsoid)} 
\eea
and, as it is well known, these two conditions determine the equilibrium 
sequence of Jacobi ellipsoids. In particular, the first represents a 
relation between the two axial ratios, while the latter 
gives the angular velocity. When $a_1 = a_2$, we obtain   
\be
\label{025} 
a_1^4  A_{11} =   a_3^2 A_3  \, , \quad \quad {\Omega^2 \over \pi \rho_0 } 
=  2 B_{11} \, , 
\ee 
or 
\be
\label{026} 
{\Omega^2 \over \pi \rho_0 } =  2 \left ( A_1 - { a_3^2 \over a_1^2 } 
A_3 \right ) \, , \quad \quad \quad \quad {\rm (Maclaurin~spheroid)}
\ee
which recovers the Maclaurin sequence. To determine the condition for 
the onset of the secular instability to nonaxisymmetric perturbations 
(LRS), we solve 
\be 
\label{027} 
{\rm det} \left ( \ders{E} {\lambda_i }{\lambda_j} \right ) _{eq} = 0 \, , 
\quad i,j = 1,2 \, .  
\ee
Since we have
\be 
\label{028} 
\left ( \derss {E} {\lambda_1} \right )_{eq} = 
\left ( \derss {E} {\lambda_2} \right )_{eq} 
\ee
the determinant vanishes at the two points 
\be 
\label{029} 
\left ( \derss {E} {\lambda_1} \right )_{eq} = 
\pm \left ( \ders{E}{ \lambda_1}{\lambda_2} \right )_{eq} \, .  
\ee
However, only the plus sign is relevant: it is easy to demonstrate 
that the minus sign corresponds to a negative value of the ratio 
$\Omega^2/(\pi \rho_0)$ and must be discarded.  
The relevant solution gives the condition
\be 
\label{029a} 
\left( \derss{I_{Ch}} {\lambda_1} \right)_{eq} + t 
\left( \derss{h} {\lambda_1} \right)_{eq}  = 
\left( \ders{I_{Ch}} {\lambda_1} {\lambda_2} \right)_{eq} + t 
\left( \ders{h} {\lambda_1} {\lambda_2} \right)_{eq}  \, . 
\ee
Finally, using the equations of the second derivatives 
reported in LRS [equations A(3), (A9)], equation (\ref{029a}) can be cast in 
the form \be \label{030} 
{\Omega^2 \over \pi \rho_0 } = 2 B_{11} \, , 
\ee
which is the exact expression for the secular instability point in
Maclaurin spheroids with respect to nonaxisymmetric perturbations (see 
Ch69, LRS). Given the constraint that the deformations be ellipsoidal, we
have identified the onset of the bar mode ($m=2$) instability point. Here 
the instability we have located would be triggered by the 
presence of viscosity, since uniform rotation was assumed in building 
the equilibrium sequence and maintained in the variations, while holding 
fixed $M_0$ and $J$ (but not circulation). 
We can also recover the result that this condition occurs at the
precise point 
where the Jacobi sequence bifurcates from the Maclaurin sequence [see 
(\ref{024}), (\ref{025})].   

Using the variational method in the framework in Newtonian physics, 
LRS were able to investigate a number of problems. In particular, they 
constructed 
approximate hydrostatic equilibrium solutions for rotating polytropes, 
either isolated or in binary systems, and presented the compressible 
generalization of the classical sequences of Maclaurin spheroids, Jacobi, 
Dedekind and Riemann ellipsoids and Roche, Darwin and Roche--Riemann 
binaries. For the case of incompressible Maclaurin sequences the 
ellipsoidal approximation is exact and the results derived from the 
variational principle are exact (Ch69). 
In this paper we will use the same variational approach 
to investigate the ellipsoidal configurations in PN gravitation. 

\section{The Post--Newtonian Solution}

We will construct a variational expression for the total mass--energy $M$ of 
a rotating ellipsoid in PN gravitation. The variation in $M$ 
is equivalent to a variation in $E = M-M_0$ for fixed $M_0$, 
and we will consider 
incompressible configurations with constant rest mass density $\rho_0$. 
For a PN treatment the evaluation of all the 
integral quantities (e.g. $M$, $M_0$ and $J$) requires a knowledge 
of the metric, which must be determined self--consistently with the 
matter profile by solving the Einstein field equations. 
The field equations 
form a set of coupled, nonlinear partial differential equations and, 
in general, must be solved numerically. 
The solution of these equations for a rotating star in general relativity 
has been tackled by many authors (see e.g. CST92; Cook, Shapiro \& 
Teukolsky 1994a,b; CST96, and references therein). 
A great simplification arises if we adopt the ``conformal approximation'' 
of CST96 (see also Wilson \& Matthews 1989, 1995; Wilson 1990, Wilson, 
Matthews \& Marronetti 1996). 
This form for the metric is only approximate (although the approximation
is very good) for rotating stars in full general relativity (CST96), for 
two main reasons. First, even in absence of 
gravitational radiation (e.g., the case for stationary, 
axisymmetric equilibrium rotating stars) the exact solution is not
conformal. 
However, as shown by CST96, the deviation 
from conformal flatness is small ($< 1 \%$ even for highly relativistic
stars). The conformal approximation is exact in PN gravitation and
departures only enter at the 2 PN order. The second 
reason is that 
emission of gravitational radiation causes the system to evolve in time, 
so the conformal decomposition of CST96 does not hold in general. However,
the 
timescale for the evolution due to gravitational radiation is much longer
than 
the dynamical timescale, and in many situations relativistic systems can be 
treated to be in quasi-equilibrium (see e.g. Baumgarte \etal 1997a,b). An 
analogous approximation is often 
used in stellar evolution calculations, where the relevant evolution 
timescales are the nuclear or Kelvin--Helmholtz timescales, so that the 
stars maintain (quasi) hydrostatic equilibrium on a dynamical timescale.
Gravitational radiation only enters at the 2.5 PN order. 
Henceforth, we 
restrict our attention to configurations in PN theory, where the 3--metric 
is conformally flat. We choose to use the 3+1 decomposition of
Einstein's equations in 
our analysis, rather than the conventional Chandrasekhar (1965a) PN 
expansion. We do so in preparation for future numerical treatments of 
nonaxisymmetric instabilities and their nonlinear evolution, which are best 
tracked in 3+1 form (see also Bonazzolla, Frieben and Gourgoulhon 1997).

Henceforth we confine our 
attention to the case of 
incompressible, rigidly rotating models, for which the PN system can be 
solved analytically provided we adopt a ellipsoidal model for the 
matter profile. 

We solve the conformal ADM equations for the metric in this section. We 
introduce our approximations, section 4.1, referring to CST92 and CST96 
for the notation and for a more detailed discussion. The final 
metric is presented in section (4.2) and expressions for the conserved 
quantities are derived in (4.3). 
Our field equations are equivalent to those of CN in the PN limit.  

\subsection{{\it Basic Equations}}  

Let us consider an isolated, self--gravitating, homogeneous system 
with rest mass density $\rho_0$ and total mass--energy density $\rho$. Assume that it is 
uniformly rotating with a constant angular velocity $\Omega$. 
In the incompressible limit, the internal 
energy is zero, and the energy-momentum tensor takes the form
\be
T_{\mu \nu  } = \left ( \rho_0 + P \right ) U_\mu U_\nu + P g_{\mu  \nu} 
\, , \ee
where $U^\mu$ is the fluid four--velocity, $g_{\mu \nu}$ are the metric 
coefficients and $P$ is the 
pressure. Here Greek indices $\mu,\nu,\dots$ range over $0 \dots  3$, 
while Latin indices $i,j,\dots$ range over $1 \dots 3$; geometrized 
units ($c \, = \, G \, = \, 1$) are used throughout. 
We assume that the outer surface 
(where $\rho = 0$) is a triaxial ellipsoid with semiaxes $a_1$, 
$a_2$, $a_3$.\footnote{Here the ellipsoid is defined in coordinate 
space, so 
that the proper shape of the PN configuration is not in general 
ellipsoidal. Although convenient mathematically, such a trial function is 
only an approximation (except in the Newtonian limit, where it is exact).} 
Following CST96, we 
start from the most general expression for the metric in a 3+1 form 
\be
ds^2 = -\alpha^2 dt^2 + \gamma_{i j} \left ( dx^i + \beta^i dt 
\right ) \left ( dx^j + \beta^j dt \right ) \, , 
\ee
where $\alpha$ and $\beta^i$ are the 
lapse and the shift functions, respectively. 
We choose a conformally flat decomposition of the spatial metric
\be
\gamma_{ij } = \Psi^4 f_{ij } \, . 
\ee
Here $\Psi $ is the conformal factor and $f_{ij}$ is the Euclidean 
metric in the adopted coordinate system. In the following we 
will use cartesian coordinates $x_i$, $i=1\dots 3$. Following CST96, we set 
$ \partial_t \left ( \gamma^{-1/3} \gamma_{ij} \right ) = 0$ and adopt the 
maximal slicing condition $ K = K_i^i = 0 $, where 
$K_{i j}$ is the extrinsic curvature. We obtain [see equation (4) by CST96] 
\be 
\label{kappa}
K_{ij } = { 1 \over 2 \alpha} \left (D_i \beta_j + D_j
\beta_i - { 2 \over 3} \gamma_{ij} D_k \beta^k\, 
\right ). \ee
Here $D_i$ indicates the covariant derivative with respect to 
$\gamma_{ij }$. 
The metric coefficients depend on the three 
functions $\alpha$, $\beta^i$ and $\Psi$ that, following CST96, 
can be derived as a 
solution of a system of ADM partial differential equations [see equations 
(8), (14) and (18) in CST96]. The first two equations are the Hamiltonian 
constraint equation and the lapse equation ($\partial_t K = 0$)  
\be 
\label{1}
\nabla^2 \Psi = -{ 1 \over 8 } \Psi^5 K^{ij} K_{ij} - 2 \pi
\Psi^5 \rho  \, , 
\ee
\be
\label{2}
\nabla^2 \left (\alpha \Psi \right ) = \left (\alpha \Psi \right )
\left [ { 7 \over 8 } \Psi ^4 K_{ij } K^{ij } + 2 \pi
\Psi^4 \left ( \rho + 2 S \right ) \right ] \, , 
\ee
while the differential equation for the shift vector can be obtained by 
substituting 
expression (\ref{kappa}) into the momentum constraint
\be 
\label{mc}
D_{i} K^{ij} = 8 \pi S^{j} \, . 
\ee
The final result is 
\bea
\label{3}
\nabla^2 \beta^i + {1 \over 3} \nabla^i \left ( \nabla_j \beta^j
\right ) &=& \left ( { 1 \over \alpha} \nabla_j \alpha - { 6 \over \Psi }
\nabla_j \Psi  \right ) 
\left ( \nabla^j \beta^i + \nabla^i 
\beta^j
- { 2 \over 3 } f^{ij} \nabla_k \beta^k \right ) \\
&+& \nonumber 16
\pi \alpha \Psi^4 S^i \, . 
\eea 
In the previous expressions $\nabla_i$ and $\nabla^2$ denote the 
flat space covariant 
derivative and the Laplacian operator.
The density $\rho$ appearing into equations (\ref{1}), 
(\ref{2}) can be derived from the stress--energy tensor  
\be
\label{4}
\rho = n^\mu n^\nu T_{\nu \mu } = \left ( \rho_0 + P \right ) \left ( \alpha 
U^t \right )^2 - P \, , 
\ee
where $n^\mu$ is the normal vector to a $t=constant$ surface.
Note that $\rho= \rho_0$ for a nonrotating sphere, since there is no
internal energy. However, 
the two densities differ in general and $\rho$ is not a 
constant, due to rotational energy contributions. The source term $S$ and
the momentum source $S^i$ are 
given by \be
\label{5}
S = \gamma^{ij} T_{ij} = \left ( \rho_0 + P \right ) \left [ 
\left ( \alpha U^t \right )^2 -1 \right ] + 3 P \, , 
\ee
\be
\label{6}
S^i
 =  -\gamma^{i}_{j} n_{k}T^{jk}
=  \left ( \rho_0 + P \right ) \left ( \alpha U^t \right ) 
\gamma^{ij } U_j \, . 
\ee

Equation (\ref{3}) can be 
conveniently reduced to 
two simpler equations by introducing the decomposition 
\be 
\label{7} 
\beta^i = G^i - { 1 \over 4 } \nabla^i B \, .   
\ee
The two equations that must be solved now become 
\be \label{8}
\nabla^2 G^i = \left ( { 1 \over \alpha} 
\nabla_j \alpha - { 6 \over \Psi }
\nabla_j \Psi  \right ) 
\left ( \nabla^j \beta^i + \nabla^i 
\beta^j
- { 2 \over 3 } f^{ij } \nabla_k \beta^k \right ) 
+16
\pi \alpha \Psi^4 S^i \, ,  
\ee
\be 
\label{9} \nabla^2 B = \nabla_k G^k \, . 
\ee 

In the fully 
relativistic case the solution of the coupled ADM equations (\ref{1}), 
(\ref{2}), 
(\ref{8}) and (\ref{9}) represents a nontrivial problem and must 
be tackled numerically. In this paper we work at the PN order and, in the 
ellipsoidal approximation, the equations can be solved analytically. 
The metric and the stress tensor will be 
expanded as sums of 
terms of successively higher order in the expansion parameter $1/c^2$, 
while each ADM equation will be decomposed into a series of equations of 
successively highly order in $1/c^2$. The first PN correction will refer
to the terms that are $O(c^{-2})$ (i.e., $O(M/R)$, where $R$ is a 
length scale of the problem) higher than the 
corresponding Newtonian terms in this expansion. 
We do not restrict our analysis to slow rotation (Hartle 1967), whereby 
one requires  $ \sqrt { R^3 \over M} \Omega \ll 1$. In that 
context, rotation is considered ``slow'' if its effects on the structure 
of the star are relatively small. Here we allow arbitrary fast rotation, 
so that $\Omega^2$ is permitted to reach $\sim \left ( M/R^3 \right )$ 
and stars can suffer considerable rotational distortion.  

To obtain equations correct to the PN order, 
in the right--hand sides 
of (\ref{1}), (\ref{2}) and (\ref{3}) we need retain 
only the contributions of order 
\be
\rho_0 \, , \quad \quad \rho_0 { M \over R} \, .   
\ee
Consider the Hamiltonian constraint and the lapse 
equations. 
From equations (\ref{3}) and (\ref{6}) we have, at the 
leading order,  
\be
S^i \sim \rho_0 v \quad  \Rightarrow \quad \beta^i \sim S^i R^2 \sim 
\rho_0 v R^2 \sim v { M \over R} \, , 
\ee
which yields [see equation (\ref{kappa})]: 
\be
K_{ij } K^{ij } \sim \left ({\beta^i \over R } \right )^2 \sim  
\rho_0 { M^2 \over R^2 }\, . 
\ee
This means that, at our order of approximation, we can safely drop 
all terms involving the extrinsic curvature from equations (\ref{1}) and 
(\ref{2}), which then reduce to the simpler form 
\be
\label{10}
\nabla^2 \Psi = - 2 \pi 
\Psi^5 \rho  \, , 
\ee
\be
\label{11}
\nabla^2 \left (\alpha \Psi \right ) =  2 \pi \alpha \Psi^5 
\left ( \rho + 2 S \right )  \, . 
\ee
We thus need approximate expressions for the source terms 
$\rho$ and $S$, and these can be obtained by expanding the product 
$(\alpha U^t)$ appearing in (\ref{4}) and (\ref{5}). The normalization 
condition 
$U_\nu U^\nu = -1$ yields 
\be 
\label{12}
(\alpha U^t )^2 = 1 + \gamma^{ij} U_iU_j \, . 
\ee
We introduce  $v^i \equiv U^i /U^t$ and consider a 
velocity field 
corresponding to uniform rotation with angular velocity $\Omega = 
U^\phi / U^t$ about the $x_3$ direction. Then 
\be 
\label{13}
v_1 = - \Omega x_2 \, , \quad
v_2 = \Omega x_1 \, , \quad
v_3 = 0 \, ,  
\ee
and $v^2 = \Omega^2 (x_1^2 + x_2^2)$. 
At the leading term, it follows that 
\be 
\label{15}
\left ( \alpha U^t \right )^2 \approx
1 + v^2 + \dots \, .  
\ee
Substituting back in the definition of $\rho$ and $S$ and recalling that 
$P/ \rho_0 \sim O (M/R) $, we 
find that, at Newtonian order, 
these two quantities can be approximated as
\be
\label{16}
\rho \approx  \rho_0 ( 1 + v^2 + \dots) \, , 
\ee
\be
\label{17}
S  \approx \rho_0  \left ( v^2 + 3 { P \over \rho_0 } + \dots \right ) 
\, . \ee
We also need the pressure distribution in the 
equilibrium configuration. Terms involving $P$ appear only in the PN 
correction. This means that we can use the Newtonian result 
for $P$ (see e.g. Chandrasekhar 1965a, hereafter Ch65a) 
\be
\label{18}
{ P \over \rho_0 } = \pi \rho_0 \left [ A_3 a_3^2 - \sum_i A_i x_i^2 
\right ] + {1 
\over 2} \Omega^2 \left ( x_1^2 + x_2^2 \right ) \, .  
\ee

The functions $\Psi$ and $\alpha$ now can be derived 
by solving the two equations 
(\ref{10}) and (\ref{11}). The conformal factor can be expanded as: 
\be
\label{19}
\Psi \equiv  1 - \Phi/2 \, , 
\ee
where $\Phi \equiv \Phi_N + \Phi_{PN}$ and where 
\be 
\label{21}
\nabla^2 \Phi_N \equiv   4 \pi  \rho_0  \, .   
\ee
Note that $\Phi_N$ represents the 
Newtonian potential and coincides with the quantity $-U$ in the 
Chandrasekhar's notation. 
Linearizing $\Psi^5 \approx \left ( 1 - 5 \Phi/2 \right) $ and 
substituting (\ref{16}), (\ref{19}), (\ref{21}) into (\ref{10}), we 
obtain  
\be
\label{22}
\nabla^2 \Phi_{PN} = - 10 \pi \rho_0 \Phi_N  + 4 \pi \rho_0 v^2 
\, . 
\ee

In a similar way, the lapse equation can be linearized by 
introducing the expansion $\alpha \Psi \equiv 1 + \Theta$, where $\Theta 
\equiv \Theta_N + \Theta_{PN}$ and $\nabla^2 \Theta_N \equiv  2 \pi
\rho_0$, so 
that $\Theta_N = \Phi_N/2$. Writing 
\be 
\label{24}
\Psi^4 \approx \left ( 1 - { \Phi_N \over 2 } \dots \right )^4 
\approx 1 - 2 \Phi_N \dots \, , 
\ee
and 
\be
\label{25}
\rho + 2 S \approx \rho_0 \left ( 1 + 3 v^2 + 6 { P \over \rho_0 } 
\dots \right ) \, , 
\ee 
we obtain 
\be 
\label{28}
\nabla^2 \Theta_{PN} = 6 \pi \rho_0 \left ( v^2 + 2 
 { P \over \rho_0 } - { 1 \over 2} \Phi_N \right ) \, . 
\ee
The corresponding expansion of the lapse function is 
\be
\label{30}
\alpha = (1 + \Theta)  \Psi^{-1} 
\approx 1+
{\Phi_N \over 2} + {\Phi_{PN} \over 2} + 
{\Phi_N^2 \over 4} + \Theta_N + \Theta_N 
{\Phi_N \over 2} 
+ \Theta_{PN} \, .  
\ee
It is straightforward to identify the Newtonian and the PN 
contributions to $\alpha$ as
\be
\label{31}
\alpha_N = \Phi_N \, , 
\ee
\be
\label{32}
\alpha_{PN} = {\Phi_N^2 \over 2} + {\Phi_{PN} \over 2} 
+ \Theta_{PN} \, . 
\ee
Parenthetically, we note that 
$\alpha_{N}$ and $\alpha_{PN}$ coincide with the correspondent terms 
derived by CN (see Appendix A.2). 
Summarizing, to derive the PN corrections 
to the lapse function and the conformal factor, we need to solve the 
system of elliptic equations (\ref{21}), (\ref{22}) and 
(\ref{28}). 

Let us now focus on the shift vector, and introduce the 
expansions
\be \label{33}
\beta^i = \beta^i_{PN} + O(c^{-5})\, , \quad 
G^i = G^i_{PN} 
+ O(c^{-5}) \, , \quad 
B = B_{PN} + O(c^{-5}) \, .  
\ee
At our order of approximation, 
we only need retain the leading terms in these expansions, of order $\sim v M/R \sim 
c^{-3}$. As a consequence, we can 
legitimately drop the first term on the 
right hand of equation (\ref{8}), being of order 
\be 
{M \over R^2} { \beta^i \over R} \, , 
\ee 
(note that $\nabla \alpha \sim \nabla \Phi \sim M /R^2$). This yields
\be \label{34}
\nabla^2 G_{PN}^i  \approx 16 \pi \alpha \Phi^4 
S^i \approx 16 \pi S^i \, , 
\ee
and we are left to derive the leading term in $S^i$. In expression 
(\ref{6}) we have 
\bea \label{35}
\gamma^{ij} U_{j} &=&    
\gamma^{ij} g_{j \nu} U^\nu \\ \nonumber 
&=& U^t \left ( \beta^i + v^i \right ) \approx v^i \, ,       
\eea
which yields to leading order 
\be \label{36}
S^i \approx \rho_0 v^i \, . 
\ee
By substituting expressions (\ref{13}) for the components of $v^i$, we 
can finally write the two equations 
\be 
\label{37}
\nabla^2 G_{PN}^1 = 
16 \pi \rho_0 v^1 = 
- 16 \pi \rho_0 \Omega x_2 \, , 
\ee 
\be 
\label{38}
\nabla^2 G_{PN}^2 = 
16 \pi \rho_0 v^2 = 
16 \pi \rho_0 \Omega x_1 \, , 
\ee 
while $G_{PN}^3 = 0$. 
Once $G_{PN}^i$ is known, $B_{PN}$ 
can be obtained simply by solving 
\be
\label{39}
\nabla^2 B_{PN} =  
\der {G_{PN}^1}{x_1} + \der {G_{PN}^2}{x_2}  \, , 
\ee
and $\beta_{PN}^i$ will follow from (\ref{7}).  

As it will be shown in the next section, the full system of elliptic 
equations (\ref{21}), (\ref{22}), (\ref{28}), (\ref{37})--(\ref{39}) 
admit an analytical solution (up to well known elliptic integrals). 

\subsection{{\it The Metric: Analytic Solution}}  

The Newtonian potential $\Phi_N$ at an internal point 
$x_i$ of a homogeneous, ellipsoidal configuration with semi--axes 
$a_i$ has the well known analytical form  
\be
\label{41}
\Phi_N 
= - \rho_0 \int  {1  
\over \left | \x - \x ' \right | } d^3 \x ' 
= - \pi \rho_0 \left (I_{Ch} - \sum_i A_i x_i^2 \right ) \, , 
\ee
[see e.g. Ch69, equation (3.40)]. This function provides the leading term in the 
expansions for $\Psi$ and $\alpha$ [equations (\ref{19}), (\ref{31})]. Hence we only 
need to determine the post--Newtonian corrections $\Phi_{PN}$, 
$\alpha_{PN}$ and $\beta^i_{PN}$ to obtain the metric at the PN order. 
The main advantage of our decomposition comes from 
the fact that the PN contributions can all be written in 
integral form by 
exploiting the Green's functions of the corresponding differential 
equations. Moreover, in our homogeneous, rigidly rotating case, the 
aforementioned integrals involve familiar quadratures of the kind \be
\rho_0 \int  {1  
\over \left | \x - \x ' \right | } d^3 \x ' \, , \qquad 
\rho_0 \int  { x_i  
\over \left | \x - \x ' \right | } d^3 \x ' \, , \qquad
\rho_0 \int  { x_i x_j 
\over \left | \x - \x ' \right | } d^3 \x ' \, .  
\ee
The first one is simply $-\Phi_N$, while the others 
coincide with the two Newtonian potentials $D_i$ and 
$D_{ij}$ given by Ch69 [see equations (3.120), 
(3.131)]. As a result, we are able to 
derive explicit expressions for the three PN corrections $\Phi_{PN}$,
$\alpha_{PN}$ and $\beta^i_{PN}$ in terms of elliptic integrals.

Consider first the PN correction to the conformal 
factor. Substituting $v^2 = \Omega^2 
\left (  
x_1^2 + x_2^2 \right )$ into the right hand of equation (\ref{22}), the 
formal solution for $\Phi_{PN}$ can be expressed as 
\be \label{42}
\Phi_{PN} (\x ) = 
{ 5 \over 2 } \rho_0 \int  {\Phi_N \left ( \x ' \right ) 
\over \left | \x - \x ' \right | } d^3 \x ' - \rho_0 \Omega^2 \int { 
\left ( x_1'^2 + x_2'^2 \right ) 
\over \left | \x - \x ' \right | } d^3 \x ' \, . 
\ee
Inserting the analytical 
expression (\ref{41}) for $\Phi_N$ in the integrand of (\ref{42}), it is
easy to recognize that $\Phi_{PN}$ can 
be written as 
\be 
\label{46}
\Phi_{PN} = { 5 \over 2 } \pi \rho_0  I_{Ch} \Phi_N + 
{ 5 \over 2 } \pi \rho_0
\sum_i A_i 
D_{ii}  - \Omega^2 \left ( D_{11} + D_{22} \right ) \, ,   
\ee
where the potential $D_{ii}$, written in terms of the index symbols 
$A_{ijk \dots}$ and $B_{ijk 
\dots}$, is [see equation (3.132) in Ch69, Appendix D]
\bea
\label{47}
D_{ii} & = & 
\rho_0 \int { x_i'^2 
\over \left | \x - \x ' \right | } d^3 \x '  \\ \nonumber 
&= & \pi \rho_0 a_i^4 \left ( A_{ii} - \sum_j A_{iij} 
x_j^2 \right ) x_i^2 
+ { 1 \over 4 } \pi \rho_0 a_i^2 \left ( B_i - 2 \sum_j B_{ij} 
x_j^2 
+ \sum_{i,k} B_{ijk} x_j^2 x_k^2 \right ) \, .  \eea 

The derivation of $\alpha_{PN}$ and $\beta_i^{PN}$ is carried out in a 
similar way. The formal solution of equation (\ref{28}) is 
\be 
\label{48}
\Theta_{PN} = 
- { 3 \over 2} \rho_0 \Omega^2 \int
{ \left ( x_1'^2 + x_2'^2 \right ) 
\over \left | \x - \x ' \right | } d^3 \x ' 
- 3 \int { P \left ( \x' \right ) \over 
\left | \x - \x ' \right | } d^3 \x ' 
+ { 3 \over 4} \rho_0 \int
{\Phi_N \left ( \x ' \right ) 
\over \left | \x - \x ' \right | } d^3 \x ' \, ,  
\ee
and, by substituting $\Phi_N$ and the pressure profile (\ref{18}), we 
obtain \be \label{49}
\Theta_{PN} = 
3 \pi \rho_0 \left ( A_3 a_3^2 + {I_{Ch} \over 4 } \right ) \Phi_N + { 15 
\over 4 } \pi \rho_0 \sum_i A_i D_{ii} - 3 \Omega^2 \left 
(D_{11} + D_{22} \right )  \, . \ee
Once 
$\Theta_{PN}$ and $\Phi_{PN}$ are known, the PN correction 
$\alpha_{PN}$ follows trivially from (\ref{32}). 

Consider next the shift vector. Solving equations (\ref{37}), (\ref{38}) for $G^1_{PN}$ 
and $G^2_{PN}$ yields 
\be  
\label{50}
G^1_{PN} = 4 \rho_0 \Omega \int 
{ x_2' 
\over \left | \x - \x ' \right | } d^3 \x ' \, , \quad 
G^2_{PN} = - 4 \rho_0 \Omega 
\int
{ x_1' 
\over \left | \x - \x ' \right | } d^3 \x ' \, ,
\ee
hence
\be  
\label{51}
G^1_{PN} = 4 \Omega D_2 \, , \quad 
G^2_{PN} = - 4 \Omega D_1 \, , \quad  
\ee
where [see equation (3.121) by Ch69] 
\bea 
\label{52}
D_i &= & \rho_0 
\int 
{ x_i' 
\over \left | \x - \x ' \right | } d^3 \x ' \\ \nonumber 
&=& \pi \rho_0 a_i^2 \left ( A_i - \sum_j A_{ij} x_j^2 
\right ) x_i \, . 
\eea 
Then, by differentiating expressions (\ref{52}) with respect to $x_i$, 
we obtain the source term appearing in the differential equation (\ref{39}) for 
$B_{PN}$. This gives 
\be 
\label{53}
\nabla^2 B_{PN} = 8 \pi \Omega \rho_0 A_{12} \left ( a_1^2 - a_2^2 \right 
) x_1 x_2 \, , \ee
where we used the symmetry property $A_{12} = 
A_{21}$. The solution is 
\bea
\label{54}
 B_{PN} 
& =&  -2 \Omega \rho_0 A_{12} \left ( a_1^2 - a_2^2 \right ) 
\int 
{ x_1' x_2' 
\over \left | \x - \x ' \right | } d^3 \x ' \\ \nonumber 
& =&  -2 \Omega A_{12} \left ( a_1^2 - a_2^2 \right ) D_{12} \, , 
\eea
where 
\be 
\label{55}
D_{12} = \pi  \rho_0 a_1^2 a_2^2 \left (A_{12} - \sum_i A_{12 i} 
x_i^2 \right ) x_1 x_2 \, . 
\ee
Note that in axisymmetry, when $a_1 = a_2$, $B = 0$ and the shift vector 
is divergence--free [see equations (\ref{7}), (\ref{9})].  
To obtain the shift from (\ref{7}), we must evaluate the 
gradient of the $D_{12}$ according to 
\be 
\label{56}
\nabla^i D_{12} = { D_{12} \over x_i} \left (\delta_{i 1} + 
\delta_{i 2} \right ) - 2 \pi \rho_0 a_1^2 a_2^2 A_{12i} x_1 x_2 x_i
\, , \ee
where $\delta_{ij}$ is the Kronecker delta. 
The final result is 
\bea 
\label{57}
\beta_{PN}^1 &=& 4 \Omega D_2 + { 1 \over 2} \Omega A_{12} { a_1^2 - 
a_2^2 \over x_1 } \left ( D_{12} - 2 \pi \rho_0 a_1^2 a_2^2 x_1^3 x_2 
A_{121} \right ) \\ \nonumber 
\beta_{PN}^2 &=& - 4 \Omega D_1 + 
{ 1 \over 2} \Omega A_{12} { a_1^2 - 
a_2^2 \over x_2 } \left ( D_{12} - 2 \pi \rho_0 a_1^2 a_2^2 x_1 x_2^3 
A_{122} \right ) \\ \nonumber 
\beta_{PN}^3 &=&  
- \Omega A_{12} \left ( a_1^2 - 
a_2^2 \right )  \pi \rho_0 a_1^2 a_2^2 x_1 x_2 x_3  
A_{123} 
\, . \eea
Equations (\ref{41}), (\ref{46}), (\ref{49}) and (\ref{57}) completely 
determine the 
metric functions to PN order in the ellipsoidal 
approximation.\footnote{\label{fo1}
In the first post--Newtonian 
approximation the orders of the 
metric coefficients that are needed are 
$O(c^{-4})$, $O(c^{-3})$ and $O(c^{-2})$ for $g_{00}$, $g_{0 \alpha}$ and 
$g_{\alpha \beta}$ respectively. However, as reported in Table 1 in CN, 
to determine the total energy without any additional assumptions we need 
$g_{\alpha \beta}$ to $O(c^{-4})$. 
An alternative method consists in evaluating the 
spatial metric $g_{ij}$ to $O(c^{-2})$: in this case 
it is still possible to derive the conserved energy, provided that 
the equations are supplemented by a
condition for isentropic flow (see 
Ch65a, Chandrasekhar 1969b, c for a detailed discussion about this 
point).} 
We compare this solution in Appendix B.1 to the exact solution for the interior 
metric in full general relativity for the case of a spherical, homogeneous
configuration.  
We verify that our results agree with the PN expansion of the exact solution. 

We have checked our results whenever possible with 
the corresponding PN expressions derived by CN, and they agree. 
A comparison is provided in Appendix A. 

\subsection{{\it The Evaluation of the Conserved Quantities $M$, $M_0$ 
and $J$}} 

\subsubsection{{\it The Integral Forms}}

With the metric coefficients derived in section (4.2) we can now derive 
explicit expressions for the total mass--energy $M$, the 
total rest--mass $M_0$ and the angular momentum $J$ of a homogeneous ellipsoid. 
These functionals will be 
then employed to determine the properties of the equilibrium configurations, 
using the energy variational method. 

First consider the total baryon rest mass $M_0$, [CST92, equation (50)]
\be
\label{62}
M_0 =  \rho_0 \int_V U^t \sqrt {-g} d^3 x 
\, ,
\ee
where $g = \det g_{\nu \mu } $ and $V$ is the volume of the ellipsoid. 
In our adopted conformal gauge  
\be
\label{64}
M_0 =  \rho_0 \int_V \left ( \alpha U^t \right ) \Psi^6 d^3 x
\, . 
\ee
In order to evaluate the PN correction to $M_0$, we need to know 
the integrand in (\ref{64}) up to order 
$M^2 /R^2$. To this order the quantity $\Psi^6$ can be 
approximated as 
\bea
\label{65}
\Psi^6 & =&  \left ( 1 - {\Phi \over 2 } \right )^6 \\ \nonumber
& \approx & 1 
- 3 \Phi + { 15 \over 4} \Phi^2 \dots \\ \nonumber 
& \approx &   
1 - 3 \Phi_N - 3 \Phi_{PN}  + {15 \over 4} \Phi_N^2  + O \left ( M^3 /R^3 
\right )  
\eea
In addition, we need a similar expansion for the quantity $\alpha U^t$ 
and this can be derived from the normalization condition (\ref{12}). 
After some algebra, we obtain 
\bea
\label{66}
\left (\alpha U^t \right )^2 
& \approx & 1 + \left (U^t \right )^2 \Psi^4 
\left [ v^2 + 2 
\left ( \beta_{PN}^2 v_2 + \beta_{PN}^1 v_1 \right ) + \dots \right ]\\ 
\nonumber 
& \approx & 1 + \left (U^t \right )^2 \left ( 1 - 2 \Phi_N + \dots \right ) 
\left [ v^2 + 2 
\left ( \beta_{PN}^2 v_2 + \beta_{PN}^1 v_1 \right ) + \dots \right ] \\ 
\nonumber & \approx & 1 + \left (U^t \right )^2 \left [ 
v^2 - 2 \Phi_N v^2  
 + 2 
\left ( \beta_{PN}^2 v_2 + \beta_{PN}^1 v_1 \right ) + \dots \right ] 
\, .  
\eea 
We can proceed via successive approximations, substituting in the right 
hand of (\ref{66}) the expansion $(U^t)^2 \approx 1 + (U^t)^2_N  + \dots$, 
where $(U^t)^2_N$ is the (unknown) Newtonian contribution to $(U^t)^2$. 
This yields
\bea
\label{67}
\left (\alpha U^t \right )^2 
& \approx & 1 + \left [ 1 + \left (U^t \right )^2_N  + \dots \right ]  
\left [ 
v^2 - 2 \Phi_N v^2  
 + 2 
\left ( \beta_{PN}^2 v_2 + \beta_{PN}^1 v_1 \right ) + \dots \right ] \\ 
\nonumber & \approx & 1 + 
v^2 - 2 \Phi_N v^2  
 + 2 
\left ( \beta_{PN}^2 v_2 + \beta_{PN}^1 v_1 \right ) + 
\left (U^t \right )^2_N v^2   + O \left ( { M^3 \over R^3} \right )    
\, . 
\eea
At Newtonian order we have 
\bea
\label{68}
\left ( U^t \right )^2 &\approx &\alpha^{-2} 
\left ( 1 + v^2 \dots 
\right ) \approx  \left ( 1 - 2 \Phi_N + \dots \right ) 
\left ( 1 + v^2 \dots 
\right ) \\ \nonumber 
& \approx& 1 - 2 \Phi_N + v^2 + O \left ( {M^2 \over R^2} \right )  
\eea 
which gives $\left ( U^t \right )^2_N =  v^2 - 2 \Phi_N$, and 
\be \label{69} 
\left (\alpha U^t \right )^2 
\approx 1 + 
v^2 - 4 \Phi_N v^2  
 + 2 \left( v_i \beta^i \right)_{PN} + 
v^4   + O \left ( { M^3 \over R^3} \right )    
\, ,  
\ee 
where $\left (v_i \beta^i \right)_{PN} = \beta_{PN}^1 v_1 
+ \beta_{PN}^2 v_2$. We obtain 
\be \label{70} 
\alpha U^t \approx 1 + 
{v^2 \over 2}  - 2 \Phi_N v^2  
 + \left (v_i \beta^i \right )_{PN} + 
{ 3 \over 8 } v^4   + O \left ( { M^3 \over R^3} \right ) 
\, . 
\ee 
The two expansions (\ref{65}) and (\ref{70}) can be used together to 
yield
\be 
\label{71} 
M_0 
\approx \rho_0 \int_V \left [ 1 + { v^2 \over 2} - 3 \Phi_N - 3 \Phi_{PN} + 
{ 15 
\over 4 } \Phi_N^2  - { 7 \over 2 } \Phi_N v^2 + { 3 \over 8 } v^4 + 
\left (v_i \beta^i \right )_{PN} \right ] d^3 x \, ,  
\ee
and this expression can be written in a more compact form noting that 
\be 
\label{72}
\int_V \Phi_{PN} d^3 x = 
\int_V \left ( v^2 \Phi_{N}  - { 5 \over 2} \Phi_N^2 \right ) d^3 x \, . 
\ee
The latter result can be verified by 
substituting equations (\ref{41}), (\ref{46}) directly and integrating; 
the answer has been checked by making use of an 
algebraic manipulator (MAPLE). As a consequence, an equivalent expression for the 
baryon rest--mass is 
\be 
\label{73} 
M_0 
\approx \rho_0 \int_V \left [ 1 + { v^2 \over 2} - 3 \Phi_N + { 45 
\over 4 } \Phi_N^2  - { 13 \over 2 } \Phi_N v^2 + { 3 \over 8 } v^4 + 
\left (v_i \beta^i \right)_{PN} \right ] d^3 x \, .   
\ee

The second conserved quantity that enters in our calculation is the total 
mass--energy $M$, which we take as the ADM mass 
(see e.g. Bowen \& York, 1980)   
\be
\label{74} 
M = - {1 \over 2 \pi } \oint_{S_\infty} \nabla^i \Psi d^2 S_i = - { 1 \over 
2 \pi} \int_V \nabla^2 \Psi d^3 x \, . 
\ee   
In equation (\ref{74}) the surface integral is over the sphere at infinity, 
while Gauss's law has been used in the 
second step. Using the Hamiltonian constraint (\ref{1}), the mass can be 
rewritten as 
\be 
\label{75}
M = I_1 + I_2 \, , 
\ee
where 
\be 
\label{75a} 
I_1 = \int \Psi^5 \rho d^3 x  \, \quad \quad I_2 =  { 1 \over 16 \pi } \int 
\Psi^5 K^{ij} K_{ij} d^3 x  \, .  
\ee
To evaluate (\ref{75a}) we note 
\be 
\label{76}
\Psi^5 = \left ( 1 - { \Phi \over 2 } \right )^5 \approx
1 - { 5 \over 2} \Phi_N 
- { 5 \over 2} \Phi_{PN} +  
{ 5 \over 2} \Phi^2_N + O \left ( { M^3 \over R^3 }\right )
\, , 
\ee
while, by making use of expressions (\ref{4}), (\ref{69}), (\ref{kappa}) 
we find
\be 
\label{77} 
\rho \approx \rho_0 \left [ 1 + v^2 + v^4 - 4 \Phi_N v^2  + 2 \left ( v_i 
\beta^i \right )_{PN} + { P \over \rho_0} v^2 \right ] 
+ O \left ( { M^3 \over R^3 }\right )
\, , 
\ee
\bea 
\label{78} \nonumber 
K_{ij} K^{ij} &\approx&  
{ 1 \over 2} \left [ 
\left ( \beta^1_{,2} \right )^2 +  
\left ( \beta_{,1}^2 \right )^2 +  
\left ( \beta_{,3}^1 \right )^2 +  
\left ( \beta_{,1}^3 \right )^2 +  
\left ( \beta_{,3}^2 \right )^2 +  
\left ( \beta_{,2}^3 \right )^2 +  
2 \beta^1_{,2} \beta^2_{,1}
+ 2 \beta^1_{,3} \beta^3_{,1}
+ 2 \beta^2_{,3} \beta^3_{,2} \right ]  \\ 
& + & \sum_{i=1}^3 \left ( \beta_{,i}^i \right)^2 
- { 1 \over 3} \left ( \sum_{i=1}^3 
\beta^i_{,i} \right )^2  
+ O \left ( { M^3 \over R^3 } \right ) 
\, , 
\eea
where $y_{,i} \equiv \partial y /\partial x_i $. For 
simplicity, the subscript ``$PN$'' has been dropped and we simply write 
$\beta^i$ for the leading term $\beta^i_{PN}$. 
The explicit form of (\ref{78}) in terms of the index symbols is tedious but 
straightforward and is obtained by differentiating 
the shift function given by (\ref{57}). The resulting 
expression has been computed with MAPLE. 
By collecting these quantities, exploiting (\ref{72}) again and 
dropping from the integrand all terms of higher order, we obtain 
\bea \nonumber 
\label{79}
M 
&\approx&  \rho_0 \int_V \left [ 1 + v^2 - { 5 \over 2} \Phi_N + { 35 
\over 4} \Phi_N^2 - 9 \Phi_N v^2 + { P \over \rho_0 } v^2 + v^4 + 2 \left 
( v_i \beta^i \right )_{PN} \right ] d^3 x \\ 
& + & {1 \over 16 \pi} \int_V K_{ij}K^{ij} d^3 x  
\, .  
\eea
For later applications it is also useful to write down the conserved energy $E = M - 
M_0$ 
\bea 
\nonumber 
\label{mmo} 
M - M_0 
&\approx&  \rho_0 \int_V \left [  
{ 1 \over 2} \Phi_N + { 1 \over 2} v^2 
- { 5 
\over 2} \Phi_N^2 - { 5 \over 2}  \Phi_N v^2 + { P \over \rho_0 } v^2 + { 
5 \over 8 } v^4 + \left ( v_i \beta^i \right )_{PN} 
\right ] d^3 x \\ & + & { 1 \over 16 \pi} \int_V K_{ij}K^{ij} d^3 x  
\, . \eea

Finally, we need to evaluate the total angular 
momentum, $J$. This quantity is obtained from the integral 
(see e.g. Bowen \& York, 1980) 
\be 
\label{80} 
J_i = { 1 \over 16 \pi} \epsilon_{ijk} \oint_{S_\infty} \left ( x^j K^{km} - 
x^k K^{jm} \right ) d^2S_m \, , 
\ee 
where $\epsilon_{ijk} $ is the Levi--Civita tensor and integration is 
over the sphere at infinity. 
Let us also introduce the symmetric tensor $ \hat K^{ij} = \Psi^2 
K^{ij}$. 
Since it is $\Psi \to 1 + O \left (r^{-1} \right )$ as $r \to \infty$, 
$  \hat K^{ij}$ may be used in place of the extrinsic curvature in 
(\ref{80}), to lowest order. Moreover, again following Bowen \& York (1980), this 
tensor satisfies 
\be
\label{81} 
\nabla_l \hat K^{kl} = \Psi^{10} 
D_l K^{kl} \, . 
\ee 
It follows that 
\bea 
\label{82} 
J_i  
&=&  { 1 \over 16 \pi} \epsilon_{ijk} \oint_{S_\infty} \left ( x^j 
\hat K^{km} - x^k \hat K^{jm} \right ) d^2S_m \\ \nonumber 
&=&  { 1 \over 8 \pi} \epsilon_{ijk} \oint_{S_\infty} x^j 
\hat K^{km} d^2 S_m \\ \nonumber 
&=&  { 1 \over 8 \pi} \epsilon_{ijk} \int x^j \nabla_l  
\hat K^{kl} d^3 x 
\, , 
\eea 
and, using (\ref{81}) and the momentum constraint 
(\ref{mc}), this can be rewritten as 
\be 
\label{83} 
J_i = \epsilon_{ijk} \int x^j S^k \Psi^{10} d^3 x \, . 
\ee 
For rotation about the $x_3$ axis, we have $J_3 = J$ and the total 
angular momentum turns out to be 
\bea 
\label{84} 
J &=& \epsilon_{3jk} \int x^j S^k \Psi^{10} d^3 x \\ \nonumber 
&= & \int \Psi^{10} \left ( x^1 S^2 - x^2 S^1 \right ) d^3 x 
\, , 
\eea 
Exploiting equations (\ref{6}), (\ref{15}), (\ref{35}) and (\ref{68}), 
we are able to derive the PN approximation of the integrand in (\ref{84}):
\bea 
\label{85} \nonumber 
S^i & = &  \rho_0 \left ( 1 + { P \over \rho_0 } \right ) \left ( \alpha 
U^t \right ) \gamma^{ij} U_j \\ \nonumber 
&\approx & \rho_0 \left ( 1 + { P \over \rho_0 }
\right ) \left ( 1 + { 1 \over 2 } v^2  \dots \right ) \left (1 + { 1 
\over 2 } v^2   - \Phi_N + \dots \right )   \left ( \beta^i_{PN} + v^i + 
\dots \right ) \\ 
& \approx & \rho_0 v^i \left ( 1 - \Phi_N + v^2 + 
{P \over \rho_0} \right ) + \rho_0 \beta^i_{PN} 
\, , 
\eea
and 
\bea
\label{86}
\Psi^{10} x^j S^i 
&\approx&  \left ( 1 - { \Phi_N \over 2 } + \dots \right 
)^{10} x^j  S^i \\ \nonumber 
&\approx& x^j \rho_0 \left [ v^i  \left ( 1 - 6 \Phi_N  + v^2 + 
{ P \over \rho_0 } 
\right ) 
+ \beta_{PN}^i \right ] \, . 
\eea
Substituting (\ref{86}) into (\ref{84}), we find 
\be 
\label{87} 
J \approx { \rho_0 \over \Omega} \int_V \left [ v^2 - 6 v^2 \Phi_N + v^4 + 
{ P \over \rho_0 } v^2 + \left ( v_i \beta^i \right )_{PN} \right ] d^3 x 
\, .  \ee 
where all terms of order $ O\left ( M^3 /R^3 \right ) $ have been neglected 
in the integrand. The kinetic energy is 
$T = \Omega J /2$ [see CST92, equation (56)], and it is easy to recognize that 
(\ref{87}) is correct at the PN level. 

The derivation of conservation laws in general relativity 
has been considered by Chandrasekhar (1969b) and CN, who 
used the symmetric energy--momentum complex of Landau--Lisfhitz to 
determine integral forms of conserved quantities at various post--Newtonian 
orders. Although CN do not present any expressions for the integrated 
conserved quantities, their results provide a useful tool 
to check the correctness of (\ref{mmo}) and (\ref{87}) by 
comparing the corresponding integrands. This comparison is presented in 
Appendix A.3. 

\subsubsection{{\it Evaluating The Conserved Integrals}}

Summarizing, integral expressions for the conserved quantities 
that we have derived are 
\bea
\label{98} \nonumber 
M 
&\approx&  \rho_0 \int_V \left [ 1 + v^2 - { 5 \over 2} \Phi_N + { 35 
\over 4} \Phi_N^2 - 9 \Phi_N v^2 + { P \over \rho_0 } v^2 + v^4 + 2 \left 
( v_i \beta^i \right )_{PN} \right ] d^3 x \\  \nonumber
& + & {1 \over 16 \pi} \int_V K_{ij}K^{ij} d^3 x  \, , \\ \nonumber
M_0 
&\approx &\rho_0 \int_V \left [ 1 + { v^2 \over 2} - 
3 \Phi_N - 3 \Phi_{PN} + { 15 
\over 4 } \Phi_N^2  - { 7 \over 2 } \Phi_N v^2 + { 3 \over 8 } v^4 + 
\left (v_i \beta^i \right )_{PN} \right ] d^3 x \, ,  
\\ \nonumber 
M - M_0 
&\approx&  \rho_0 \int_V \left [  
{ 1 \over 2} \Phi_N + { 1 \over 2} v^2 
- { 5 
\over 2} \Phi_N^2 - { 5 \over 2}  \Phi_N v^2 + { P \over \rho_0 } v^2 + { 
5 \over 8 } v^4 + \left ( v_i \beta^i \right )_{PN} 
\right ] d^3 x \\ \nonumber & + & { 1 \over 16 \pi} \int_V K_{ij}K^{ij} 
d^3 x  \, , \\ 
J & \approx &  { \rho_0 \over \Omega} \int_V \left [ v^2 - 6 v^2 \Phi_N + 
v^4 + 
{ P \over \rho_0 } v^2 + \left ( v_i \beta^i \right )_{PN} \right ] d^3 x 
\, , 
\eea
In order to perform the quadratures over the fluid volume, we adopt the 
ellipsoidal approximation for the matter distribution, whereas $\rho_0 = 
$ constant inside a triaxial ellipsoid. With this assumption, it is 
convenient to introduce and evaluate the following integrals:
\bea
\label{99}
{\cal I}_1 &=& \int_V \rho_0 d^3 x \equiv M_c = { 4 \pi \over 3} \rho_0 a_1 
a_2 a_3  = 
{ 4 \pi \over 3} \rho_0 R^3  \, ,  
\\
{\cal I}_2 &=& \int_V \rho_0 \Phi_N d^3 x = 2 {\cal M} \equiv 
2 \sum_i {\cal M}_{ii} = - { 6 \over 5} {M_c^2 \over R } f 
\, , \\
{\cal I}_3 &=& \int_V \rho_0 v^2 d^3 x = \Omega^2 \int_V \rho_0 \left (x_1^2 
+ x_2^2 \right ) d^3 x = \Omega^2 \left (I_{11} + I_{22} \right ) = 
{ 2 \over 5 } M_c \Omega^2 R^2 {1 \over h} 
 \, , 
\eea
where ${\cal M} $ is the Newtonian potential energy, while $R$ and the two 
dimensionless ratios $f$ 
and $h$ have been introduced in section (3) [see equations (\ref{06}), 
(\ref{015}), (\ref{016})]. In deriving the previous expressions we made 
use of the results
\bea \label{100} 
{\cal M}  &=&  \sum_i {\cal M}_{ii} \, , \\ 
{\cal M}_{ii} &=& - 2 \pi \rho_0 A_i I_{ii} \, , \\ 
I_{ij} &=& \int_V \rho_0 x_i x_j d^3 x = { 1 \over 5 } M_c a_i ^2 
\delta_{ij} \, , \eea  
given by expression (2.4), (2.12), (2.13), (3.128), (3.129) in Ch69. Note 
that, in contrast to the Newtonian analysis, 
$M_c$ is only a coordinate quantity and has no physical meaning. 
The other integrals appearing in equations (\ref{98}) are 
\bea 
\label{101}
{\cal I}_4& =& \int_V \rho_0 \Phi_{PN} d^3 x\, , \quad \quad 
{\cal I}_5 = \int_V \rho_0 \Phi_{N}^2 d^3 x\, , 
\\ \nonumber 
{\cal I}_6 &=& \int_V P v^2  d^3 x \, , \quad \quad 
{\cal I}_7 = \int_V \rho_0 \Phi_{N} v^2  d^3 x\, , 
\\ \nonumber 
{\cal I}_8 &=& \int_V \rho_0 v^4  d^3 x \, , \quad \quad 
{\cal I}_9 = \int_V \rho_0 \Omega \left ( x_1 \beta_{PN}^2 - 
\beta_{PN}^1 x_2 \right ) d^3 x\, , \\ \nonumber 
{\cal I}_{10} &=& { 1 \over 16 \pi} \int_V K^{ij}K_{ij} d^3 x \, . 
\eea 
From the quadratic forms of $\Phi_N$, $\Phi_{PN}$, etc... it is easy to 
recognize that the evaluation of ${\cal I}_i$, $i=4..9$, only involves 
quadratures of the kind $I_{ij}$ and $E_{ij}$, where\footnote{The 
integrals $E_{ij}$ are easily evaluated by 
introducing the variables $x_i' = 
x_i/a_i$, such that the equation of the ellipsoid is $\sum_i \left (x_i' 
\right )^2= 1$, and using spherical polar coordinates in the new system.}  
\be 
\label{102}
E_{ij} = \int_V \rho_0 x_i^2 x_j^2 d^3 x  = { 1 
\over 35} M_c 
a_i^2 a_j^2 \left ( 1 + 
2 \delta_{ij} \right ) 
\, ;
\ee
the same finding holds also in the case of $I_{10}$, but the integration 
involves some lengthy algebraic calculations. After many simplifications, 
and exploiting the properties of the index 
symbols, we obtain with the help of MAPLE the following integrated forms of 
the conserved quantities  \bea
\label{103a}
M & \approx & M_c + 3 
{ M_c^2 \over R} f + { 2 \over 5 } M_c 
\Omega^2 R^2 { 1 \over h} + { M_c^3 \over R^2 } g_1 + { M_c^2 \over R } 
\Omega^2 R^2 p_1 \, , \\
\label{103b}
M_0 &\approx & M_c + { 18 \over 5} { M_c^2 \over R} f + { 1 \over 5 } M_c 
\Omega^2 R^2 { 1 \over h} + { M_c^3 \over R^2 } g_2 + { M_c^2 \over R } 
\Omega^2 R^2 p_2 \, , \\
\label{103c}
J & \approx & \Omega  M_c R^2 { 2 \over 5 h }  
\left (1   + { 5 \over 2} { M_c \over R }  p_3 h \right ) \, , \\
\label{103d}
M - M_0 & \approx & - { 3 \over 5} 
{ M_c^2 \over R} f + { 1 \over 5 } M_c 
\Omega^2 R^2 { 1 \over h} + { M_c^3 \over R^2 } g_{12} + 
{ M_c^2 \over R } 
\Omega^2 R^2 p_{12} \, , 
\eea
where the functions $g_i$, $p_i$ are defined as (see Appendix D)
\bea \label{104}
g_1 & = & { 99 \over 8 } f^2 + { 9 \over 32} { 1 \over \lambda_1 
\lambda_2 } \sum_l A_l^2 {a_l^4 \over a_1^2 a_2^2 } \, , \\ 
g_2 & = & { 891 \over 56 } f^2 + { 81 \over 224} { 1 \over \lambda_1 
\lambda_2 } \sum_l A_l^2 {a_l^4 \over a_1^2 a_2^2 } \, , \\ 
g_{12} & \equiv & g_1 - g_2 = -{ 99 \over 28 } f^2 - { 9 \over 112} { 1 
\over \lambda_1 \lambda_2 } \sum_l A_l^2 {a_l^4 \over a_1^2 a_2^2 } \, , \\ 
p_1 & = & 
{ 6 \over 7} { f \over h} + { 57 \over 35 } A_3 { \lambda_1 
\lambda_2 \over h} + { 69 \over 70} { A_1 + A_2 \over \lambda_1 \lambda_2 
} + { 3 \over 70} { 1 \over \lambda_1 \lambda_2 } \left ( A_1 - A_2 
\right )^2 
+ \tilde {\cal I}_{10} \, , \\
p_2 & = & 
{ 21 \over 20} { f \over h} + { 57 \over 56 } A_3 { \lambda_1 
\lambda_2 \over h} + { 33 \over 56} { A_1 + A_2 \over \lambda_1 \lambda_2 
} + { 3 \over 140} { 1 \over \lambda_1 \lambda_2 } \left ( A_1 - A_2 
\right )^2 \, , \\
p_{12} & \equiv & p_1 - p_2 = 
-{ 27 \over 140} { f \over h} + { 171 \over 280 } A_3 { \lambda_1 
\lambda_2 \over h} + { 111 \over 280} { A_1 + A_2 \over \lambda_1 \lambda_2 
} + { 3 \over 140} { 1 \over \lambda_1 \lambda_2 } \left ( A_1 - A_2 
\right )^2  
\\ 
\label{104app}
\nonumber 
&+& \tilde {\cal I}_{10} \, , \\
p_3 &= &
{ 6 \over 5} { f \over h} + { 24 \over 35 } A_3 { \lambda_1 
\lambda_2 \over h} + { 18 \over 35} { A_1 + A_2 \over \lambda_1 \lambda_2 
} + { 3 \over 140} { 1 \over \lambda_1 \lambda_2 } \left ( A_1 - A_2 
\right )^2 \, . 
\eea   
Due to the complexity of the $K^{ij}K_{ij}$ 
contribution appearing into $M$, we simplify the notation by leaving 
the term $\tilde {\cal I}_{10} = \left ({ M_c^2 \over 
R } \Omega^2 R^2 \right )^{-1} {\cal I}_{10}$, without substituting 
the explicit form in terms of the ellipsoidal variables. For an 
expression in terms of these variables, see Appendix C. 
The structure of expressions (\ref{103a})--(\ref{103d}) is 
particularly convenient for performing the required variations, since 
the full dependence on the two axial ratios 
is contained in $f$, $h$ and, for the PN contributions, in $g_i$ 
and $p_i$. We have checked that our result agrees with the PN correction 
to the Newtonian energy obtained by ST for 
homogeneous spheres (Appendix B.2).  

%
%

\section{Equilibrium Configurations}

As we did in the Newtonian case (see 
section 3), we now construct sequences of axisymmetric equilibrium 
models. Each sequence can be parametrized by, e.g., the value 
$M / R_s $ of the spherical, nonrotating member. 
These configurations yield a PN generalization of 
Maclaurin spheroids and were originally investigated by 
Chandrasekhar (1965b) by using the tensor virial method. 
As we will confirm, the effect of 
general relativity is to attribute to a star of a given eccentricity a 
larger value of $\Omega^2 / \left ( \pi \rho_0 \right )$. Sequences of 
relativistic, numerical models at fixed $M_0 \sqrt{\rho_0}$ 
have been also published by BI (see also 
Bonazzola \& Schneider 1974). 

We construct the equilibrium sequence by minimizing 
$M$ or, equivalently, $M-M_0$, keeping fixed $M_0$ and $J$. Due to the 
complexity of the expressions, part of the calculations here 
and in the next section have been done 
by using an algebraic manipulator (MAPLE). The 
procedure is as follows. First, we combine expressions (\ref{103c}) and 
(\ref{103d}) and rewrite the quantity $M - M_0$ as a function of 
$R$, $\lambda_1$, $\lambda_2$, $\rho_0$ and $J$. This gives 
\be  
\label{113}
M - M_0 \approx  - { 3 \over 5} 
{ M_c^2 \over R} f + { 5 \over 4 } 
{ J^2 \over M_c R^2 } h  
+ { M_c^3 \over R^2 } g_{12} + 
{ 25 \over 4 } 
{ J^2 \over R^3 } h^2 p_{123} 
\, , 
\ee 
where $M_c = M_c \left( \rho_0 , R \right )$, $p_{123} = p_{12} - p_3$ 
and where we used the relation 
\bea 
\label{114}
\Omega^2 R^2 & \approx & 
{  J^2 \over M^2_c R^2 }
{ 25 h^2 \over 4} \left ( 1 + { 5  \over 2} { M_c \over R} p_3 h \right )^{-2}
\\ \nonumber 
&\approx& {  J^2 \over M^2_c R^2 }
{ 25 h^2 \over 4} \left ( 1 - 5 { M_c \over R} p_3 h \right )
\, . 
\eea
The equilibrium sequence is then determined by minimizing $M-M_0$ 
according to 
\be 
\label{115}
\der{ \left (M - M_0 \right )  }{ \lambda_i } = 
\der{ \left ( M - M_0 \right ) }{ R } 
\der{ R }{ \lambda_i } + \left [ 
\der{ \left ( M - M_0 \right ) }{ \lambda_i } \right ]_{R = const} = 0
\, , 
\ee
where, here and in the following, the partial derivative with respect to 
$\lambda_i$ is taken holding constant $J$, $\rho_0$ and $\lambda_j$, with 
$i \neq j$ 
\be \label {der}
\der{ }{ \lambda_i } \equiv 
\left ( \der{ }{ \lambda_i } \right )_{J, \rho_0, \lambda_j \left ( i 
\neq j \right )} \, . 
\ee
Note that, in contrast to the Newtonian case, the mean radius $R$ is no 
longer constant. The variation of $R$ 
can be obtained exploiting the constraint $d M_0 = 0$, which gives 
\be 
\label{116}
\der{ R }{ \lambda_i }  = - 
\left ( \der{ M_0  }{ R }  \right )^{-1} 
\left ( \der{ M_0  }{ \lambda_i }  \right )_{R = const}  \, .  
\ee 
Since 
\be \label{116a} 
M_0 \approx M_c + { 18 \over 5} { M_c \over R^2 } f + { 5 \over 4} {J^2 
\over M_c R^2} h + \dots
\, , 
\ee
expression (\ref{116}) can be approximated as 
\be 
\label{116b} 
\der{ R }{ \lambda_i }  \approx M_c  \left [
- { 6 \over 5} 
\left ( \der{ f }{ \lambda_i }  \right )_{R = const}  
- { 5 \over 12 } { J^2 \over M^3_c R} 
\left ( \der{ h }{ \lambda_i }  \right )_{R = const} \right ] 
\, . 
\ee

In axisymmetry, we set $a_1 = a_2$ when performing the two variations 
(\ref{115}). Since, in this 
case, the first derivatives 
with respect to the two axial ratios are equal, we need only consider 
the case $i=1$. To simplify the notation, we introduce the 
following symbols for the derivatives of a given function $X$
\bea 
\label{117} \nonumber 
 X^\un \equiv 
\left ( \der{ X }{ \lambda_1 }  \right )_{R = const}  
\, , \quad \quad 
 X^\du \equiv \left ( \der{ X }{ \lambda_2 }  \right )_{R = const}  
\, , \\ \nonumber
 X^\ud \equiv \left ( \ders{ X }
{ \lambda_1 } { \lambda_2 }  \right )_{R = const}  
\, \quad \quad 
 X^\uu \equiv \left ( \derss{ X }{ \lambda_1 }  \right )_{R = const}  
\, . 
\eea  
At the PN order, the variation (\ref{115}) with respect 
to $\lambda_1$ is
\be 
\label{118}
\left ( 
- 3 { M_c ^2 \over R^2 } f - 
{25 \over 4} {J^2 \over M_c R^3 } h  + \dots \right ) 
\der{R} {\lambda_1} 
- {3 \over 5} { M_c ^2 \over R } f^\un + 
{5 \over 4} {J^2 \over M_c R^2 } h^\un  
+  { M_c ^3 \over R^2 } g_{12}^\un
+ {25 \over 4} {J^2 \over R^3 } \left( h^2 p_{123} \right )^\un = 0 \, ,  
\ee
and this expression can be combined with (\ref{116b}) to yield
\bea
\label{119}
- {3 \over 5} f^\un + 
{5 \over 4} {J^2 \over M_c^3 R } h^\un   
&+& { M_c \over R} \left [ g_{12}^\un  
+ {25 \over 4} 
{J^2 \over M_c^3 R } 
\left( h^2 p_{123} \right )^\un 
+ { 18 \over 5} f f^\un 
\right .  \\ \nonumber 
& + & \left .
{ 5 \over 4 } 
{J^2 \over M_c^3 R } 
f h^\un 
+ { 15 \over 2 } 
{J^2 \over M_c^3 R } 
h f^\un 
+ { 125 \over 48 } 
{J^4 \over M_c^6 R^2 } 
h h^\un \right ] = 0 
\, . 
\eea
It is now more convenient reexpress the previous equation in terms of the 
gauge invariant parameter $\Omega^2 / \left ( \pi \rho_0 \right )$. 
From equation (\ref{103c}) we obtain 
\be 
\label{120}
{ J^2 \over M_c^3 R} \approx { \Omega^2 \over \pi \rho_0 } { 3 \over 25 
h^2 } \left ( 1 + 5 { M_c \over R} p_3 h \right ) \, . 
\ee
By using this result, the equilibrium condition (\ref{119}) becomes 
\bea 
\label{121}
{ \Omega^2 \over \pi \rho_0 } = { 4 h^2 f^\un  \over h^\un } 
& - &{ 20 \over 
3} { h^2 \over h^\un} {M_c \over R} \left 
\{ g_{12}^\un + { 18 \over 5} 
f f^\un \right . \\ \nonumber 
& + & \left . { \Omega^2 \over \pi \rho_0 } \left [ { 3 \over 4 h^2 } 
\left ( h^2 p_{123} \right )^\un + { 3 \over 4 } { h^\un \over h} p_3 + 
{ 3 \over 20  } { f h^\un \over h^2 } + {21 \over 20} {f^\un \over h} 
\right ] \right \} \, , 
\eea
where, in the right hand side, the first term represents the 
Newtonian result, while the remaining terms give the PN correction. 
Finally we set $a_1 = a_2$, using expressions (\ref{oveas1}), 
(\ref{oveas2}). The derivatives of $f$ and $h$ are derived using 
expressions (A3), (A9) in LRS, while for the derivates of $g_{12}$ 
and $p_{123}$ we 
used results reported in Appendix D.2 (expressions \ref{dpri}). 
Doing this, we obtain 
the relation between the angular velocity and the 
eccentricity along the relativistic equilibrium sequence. By adopting the 
same notation as in Chandrasekhar (1965b), we write the resulting 
expression in the form 
\be 
\label{122}
{ \Omega^2 \over \pi \rho_0 } = 8 \lambda^2 \left . f^\un \right 
|_{a_1 = a_2 } +  { 2M_c \over a_1} E \left (e \right )  
\, , 
\ee
where $\lambda \equiv \lambda_1 = \lambda_2 = \left(1 - e^2 
\right)^{1/3}$ and where the 
strength of relativity is measured by the compaction parameter $2 M_c /
a_1$, 
\bea
\label{123} 
E \left (e \right ) &=& - { 20 \over 3} \lambda^ {3/2} \left \{ 
g_{12}^\un + { 18 \over 5} 
f f^\un 
\right . \\ \nonumber 
& + & \left . 
{ \Omega^2 \over \pi \rho_0 } \left [ { 3 \over 4 
\lambda^2 } 
\left ( h^2 p_{123} \right )^\un + { 3 \over 8 } { p_3 \over \lambda} + 
{ 3 \over 40  } { f \over \lambda^2 } + {21 \over 20} {f^\un \over \lambda} 
\right ] \right \}_{a_1 = a_2 }  \, , 
\eea
is a function only of the eccentricity of the spheroid, and 
\be 
\label{124} 
\left . 8 \lambda^2 f^\un \right |_{a_1 = a_2} 
= 2 \left ( A_1 - {a_3^2 
\over a_1^2 } A_3 \right ) \, , \ee
which formally coincides with the Newtonian expression (\ref{026}). 
However, we emphasize that the eccentricity which enters in our PN 
formalism (and that of Chandrasekhar 1965b) is defined in terms of the 
ratio of the coordinate quantities 
$a_3$ and $a_1$ and is not a gauge invariant parameter. 

\placefigure{fig1}
\placefigure{fig2}

The PN sequences of equilibrium are reported in Figure \ref{fig1} for 
different values of the parameter $M/R_s$, where $R_s$ is the equatorial 
radius in Schwarzschild coordinates, and, in the spherical limit 
\be \label {gr} 
{M_c \over a_1} = { M \over R_s} { 1 \over 4} \left ( 1 - { M \over 
R_s} + \sqrt{ 1 - { 2 M \over R_s}} \right)^2 \, , 
\ee
[see Lightman \etal 1979, expression (4), page 422]. Note that this 
function has a 
maximum for $M/R_s = \left (M/R_s \right)_{max}= 5/18$, corresponding to 
$2M_c/a_1 = 3125/11664 \approx 0.268$. It 
follows that our PN formalism can be used to investigate relativistic sequences up 
to a maximum value $\left (M/R_s \right)_{max}= 5/18 \approx 0.28$. 
As we can see from Figure \ref{fig1}, we find that in the relativistic case 
the value of 
$\Omega^2 /(\pi \rho_0)$ is larger than what Newtonian theory predicts for 
the same value 
of the eccentricity, confirming the Chandrasekhar's (1965b) results. 
Sequences of equilibrium can be described in an equivalent way by using, in place of 
$\Omega^2 / (\pi \rho_0) $, the ratio 
\be 
\label{twd} 
{ T \over | W | } \equiv 
{ { 1 \over 2 } \Omega J \over \left ( { 1 \over 2 } \Omega J  + 
M_0 - M \right ) }\, , 
\ee 
which reduces to the ratio of rotational kinetic energy to gravitational potential 
energy in the Newtonian limit. Note that $T/|W|$ is gauge invariant for rigidly rotating 
objects, since $M-M_0$, $J$ and $\Omega$ are gauge invariant (they can all
be measured by observers at large distance from the spheroid). By using the
expressions of $J$, $M-M_0$ and setting $a_1 
= a_2$, we find, to PN order, 
\be
\label{tw} 
{ T \over | W | } = { \Omega^2 \over \pi \rho_0 } { 1 \over 2 \left ( 2 A_1 + A_3 
\lambda^3 \right )  } \left \{ 1 + { 2 M_c 
\over a_1 } { 5 \over 2 \sqrt{\lambda} } \left [ { p_3 h  \over 2 } + { g_{12} \over 3 f } 
+ { 1 \over 4 f } { \Omega^2 \over \pi \rho_0 } \left ( p_{12} - { 1 \over 2} p_3 \right 
) \right ] \right \}_{\lambda_1 = \lambda_2 = \lambda} \, .  
\ee 
Figure \ref{fig2} shows the sequence of equilibrium corresponding to the 
maximum value $\left (M/R_s \right)_{max}= 5/18$ together 
with the Newtonian Maclaurin sequence ($M/R_s = 0$). As we can see, in the
range we are considering, $T/|W|$ is 
insensitive to the parameter $M/R_s$. Figure \ref{fig3} shows the relation 
between the two gauge 
invariant quantities $\Omega^2 / \pi \rho_0$ and $T / |W|$, for the same equilibrium 
sequences as in Figure \ref{fig1}; squares mark the secular instability 
point and will be discussed in the next section. 
\placefigure{fig3}

\placetable{tab1}
\placefigure{fig4}

The function $E(e)$ and the corresponding function $E_{Ch}$ 
obtained by Chandrasekhar (1965b) are compared in Table \ref{tab1} and 
Figure \ref{fig4}. As it can be 
seen, the two expressions are in close agreement up at least to the value
of $e$ 
corresponding to the secular instability point in Newtonian theory, 
with a maximum 
fractional difference of a few percent. The discrepancy increases at larger 
eccentricity (30 $\%$ at the Newtonian dynamical instability point). 
This discrepancy 
may be ascribed mainly to our coordinate 
ellipsoidal approximation for the deformation and equilibrium shape. 
This assumption is only exact in the Newtonian limit. For $e \ll 1$, 
the two functions can be expanded as \be 
\label{125} 
E \approx 0.219e^2 + 0.115e^4 + 0.059e^6 + \dots \, , 
\ee
\be 
\label{126} 
E_{Ch} \approx 0.228e^2 + 0.114e^4 + 0.042e^6 + \dots \, . 
\ee

We have compared our results with the fully relativistic, numerical sequences 
presented in Figure 1 in the paper by BI. 
To compare those computations with our PN models, we have restricted
our attention to the least 
relativistic sequence in BI, which corresponds to 
$2 M_c/a_1 = 0.206$ ($\gamma_s = 0.154$ in BI). To make the 
comparison, we note that BI parametrize in 
terms of a proper eccentricity, which is defined as  
\be
\label{127} 
e_{BI}^2 = 1 - { d_p^2 \over d_e^2 }  \, , 
\ee
\be
\label{128} 
d_p = \int_0^{a_3} \left . \Psi^2 \right |_{x_1=x_2=0} d x_3 \, , 
\quad \quad d_e = \int_0^{a_1} \left . \Psi^2 \right |_{x_2=x_3=0} d x_1 
\, ,  \ee
where $d_p$ and $d_e$ are the proper radii in the polar and equatorial directions, 
respectively. 
Using (\ref{19}) and (\ref{41}) for $\Psi$, we find 
\bea
\label{129} 
e_{BI}^2 
&\approx &e^2 + { 2 M_c  \over a_1} \sqrt{1 - e^2 } \left [ { A_1 
\over 2 } \left ( {3 \over 2}  - e^2 \right ) - { 1 - e^2 \over 2} 
 \right ] \\ \nonumber 
&\approx &e^2 + { 2 M_c  \over a_1} \left ( 
{ 1 \over 15} e^2 
- { 1 \over 42} e^4 
- { 11 \over 840} e^6 + \dots \right )  
\, ,  
\eea
which provides the relation between the two definitions of eccentricity. 

\placetable{tab2}
\placefigure{fig5}
Figure \ref{fig5} shows the 
comparison between the BI values of 
$\Omega^2 / \left (\pi \rho_0 \right ) $ and the two PN sequences derived by using 
$E(e)$ and the Chandrasekhar (1965b) expression $E_{Ch}$, with $2 M_c/a_1 = 
0.206$; Table \ref{tab2} contains the values of 
$\Omega^2 / \left (\pi \rho_0 \right ) $, evaluated by using our 
expression, and the corresponding quantities by BI. 
In order to make a comparison between the relativistic corrections, which is more 
useful, we define the correction 
as the difference between the relativistic and the Newtonian 
value of $\Omega^2 / \left (\pi \rho_0 \right ) $ [given by expression 
(\ref{124})] and compare these corrections. Since BI does not present 
tabulated values for 
$\Omega^2 / \left (\pi \rho_0 \right ) $, the values we have reported in 
table \ref{tab2} have been read off Figure 1 in BI, with an estimated 
error of 
$\sim 0.02$. For $e< 0.7$, this error is of the same order as the PN 
correction, making the measured value of the latter questionable. For 
this reason we have reported in Table \ref{tab2} only the values of this 
quantity for $e$ higher than 0.7, 
together with the corresponding fractional error. 
As it can be seen, in this case our values are lower with 
respect to the numerical ones by BI, but the difference in the 
corrections is at most $\approx 10 \% $ for $e \approx 0.8$. 

\section{The Secular Instability Point}  

The determination of the point of onset of secular instability proceeds as
in the Newtonian case [cf. equations (\ref{027}), (\ref{029})], whereby we 
must evaluate
\be 
\label{130} 
\left . \derss{\left ( M - M_0 \right ) }{\lambda_1} \right |_{\lambda_1 
= \lambda_2 = \lambda} = 
\left . \ders{\left ( M - M_0 \right ) }
{\lambda_2} {\lambda_1} \right |_{\lambda_1
= \lambda_2 = \lambda} 
 \, .  
\ee
By using expression (\ref{115}), the second derivatives can be written as 
\bea 
\label{131} 
\derss{\left ( M - M_0 \right ) }{\lambda_1} & = &
\der{ \, } { \lambda_1} \left \{  
\der{ \left ( M - M_0 \right ) }{ R } 
\der{ R }{ \lambda_1 } + \left [ 
\der{ \left ( M - M_0 \right ) }{ \lambda_i } \right ]_{R = const} \right 
\} 
\, , \\ 
\label{132} 
\ders{\left ( M - M_0 \right ) }
{\lambda_2} {\lambda_1} &=&  
\der{ \, } { \lambda_2} \left \{  
\der{ \left ( M - M_0 \right ) }{ R } 
\der{ R }{ \lambda_1 } + \left [ 
\der{ \left ( M - M_0 \right ) }{ \lambda_i } \right ]_{R = const} \right 
\} 
 \, ,  
\eea
where the expression in brackets is given by 
equation (\ref{119}), 
multiplied by $M_c^2/R$. After some algebra we find that, at PN order, 
\bea 
\label{133} \nonumber 
\derss{\left ( M - M_0 \right ) }{\lambda_1} & = &
{ M_c^2 \over R } 
\left [ 
- {3 \over 5} f^\uu + 
{5 \over 4} {J^2 \over M_c^3 R } h^\uu   
\right ] 
+ { M_c^3 \over R^2 } \left [  
{ 36 \over 5} 
\left ( f^\un \right )^2 +  
{ 18 \over 5} f f^\uu 
+ { 35 \over 2} { J^2 \over M_c^3 R} f^\un h^\un  
\right . \\ & +& \left .  
{ 5 \over 4} { J^2 \over M_c^3 R} f  h^\uu 
+ {15 \over 2}  { J^2 \over M_c^3 R} h  f^\uu 
+ { 125 \over 24} 
{ J^4 \over M_c^6 R^2 } \left ( h^\un \right )^2  
+ { 125 \over 48} 
{ J^4 \over M_c^6 R^2 } h h^\uu 
\right . \\ \nonumber & +& \left .  
g_{12}^\uu + { 25 \over 4} 
 { J^2 \over M_c^3 R} \left ( h^2 p_{123} \right )^\uu
\right ]
\, , \eea
\bea 
\label{134} \nonumber 
\ders{\left ( M - M_0 \right ) }{\lambda_1} {\lambda_2} & = &
{ M_c^2 \over R } 
\left [ 
- {3 \over 5} f^\ud + 
{5 \over 4} {J^2 \over M_c^3 R } h^\ud  
\right ] 
+ { M_c^3 \over R^2 } \left [  
{ 36 \over 5} f^\du f^\un  +  
{ 18 \over 5} f f^\ud 
+ { 35 \over 4} { J^2 \over M_c^3 R} \left ( f^\un h^\du 
\right . \right . \\ \nonumber & +& \left . \left .  
f^\du h^\un \right )   +
{ 5 \over 4} { J^2 \over M_c^3 R} f  h^\ud 
+ {15 \over 2}  { J^2 \over M_c^3 R} h  f^\ud 
+ { 125 \over 24} 
{ J^4 \over M_c^6 R^2 } h^\un h^\du   
+ { 125 \over 48} 
{ J^4 \over M_c^6 R^2 } h h^\ud 
\right . \\ & +& \left .  
g_{12}^\ud + { 25 \over 4} 
 { J^2 \over M_c^3 R} \left ( h^2 p_{123} \right )^\ud  
\right ] \, .
\eea
Given axisymmetry, we can set 
$a_1 = a_2$ when equating expressions (\ref{133}) and (\ref{134}); the 
equivalence between the subscripts 
1 and 2 will be made explicit at the end of the calculations. 
After this substitution is made all terms 
containing first derivatives cancel in the final relation; hence, for the
sake of simplicity, they will be dropped in the following expressions. 
Imposing condition (\ref{130}) yields 
\bea 
\label{135}
{ 5 \over 4 } { J^2 \over M_c^3 R} \cd \left [ h \right ] & = & 
{ 3 \over 5 } \cd \left [ f \right ] 
- { M_c \over R} 
\left \{ 
{ 18 \over 5 } f \cd \left [ f \right ] 
+ { 5 \over 4 } 
{ J^2 \over M_c^3 R}
f \cd \left [ h \right ] 
+ { 15 \over 2 } 
{ J^2 \over M_c^3 R}
h \cd \left [ f \right ] \right . \\ \nonumber 
& + & \left . 
{ 125 \over 48 } 
{ J^4 \over M_c^6 R^2} h 
\cd \left [ h \right ] +
\cd \left [ g_{12} \right ] +
{ 25 \over 4 } 
{ J^2 \over M_c^3 R}
\cd \left [ h^2 p_{123} \right ] \right \} \, , 
\eea
where the difference operator $\cd$ is defined as the 
difference between the two second derivatives of a function with respect to 
the axial ratios 
\be 
\label{136}
\cd \left [ X \right ] \equiv X^\ud - X^\uu
\, . 
\ee
Replacing $J^2$ with $\Omega^2$ according to (\ref{120}), 
expression (\ref{135}) becomes 
\bea
\label{137} 
{\Omega^2 \over \pi \rho_0 } &=&  4 h^2 { \cd \left [ f \right ] \over 
\cd \left [h \right ] } + { 20 \over 3 } {M_c \over R} {h^2 \over 
\cd \left [h \right ]} 
\left \{ 
\cd \left [g_{12} \right ] 
+ { 21 \over 5} f  
\cd \left [f  \right ] 
\right . \\ \nonumber & + & \left . 
3 p_3 h 
\cd \left [f  \right ] 
 + { \Omega^2 \over \pi \rho_0 } \left [ 
{ 21 \over 20 h } 
\cd \left [f  \right ] 
+ { 3 \over 4 h^2 } 
\cd \left [ h^2 p_{123}  \right ] \right ] \right \} \, . 
\eea 
Setting now $\lambda_1 = \lambda_2 = \lambda $ gives 
\be
\label{138} 
{\Omega^2 \over \pi \rho_0 } =  { 1 \over 5} \left ( 9 B_{11} + 
A_1 - \lambda^3 A_3 \right )  + { 2 M_c  \over a_1} { 9 \over 10} P \left( e 
\right ) \, , 
\ee
where 
\bea
\label{139}
P \left (e \right ) & = & - { 20 
\over 27} \lambda^{5/2} 
\left \{ 
\cd \left [g_{12} \right ] 
+ { 21 \over 5} f  
\cd \left [f  \right ] 
\right . \\ \nonumber & + & \left . 
3 p_3 h 
\cd \left [f  \right ] 
 + { \Omega^2 \over \pi \rho_0 } \left [ 
{ 21 \over 20 h } 
\cd \left [f  \right ] 
+ { 3 \over 4 h^2 } 
\cd \left [ h^2 p_{123}  \right ] \right ] \right \}_{\lambda_1 = \lambda2} 
\,   \eea 
is a function of the eccentricity. 
The last step consists in evaluating this condition along the equilibrium 
sequence. Using equations (\ref{122}) and (\ref{124}) we obtain  
\be
\label{140}
{\Omega^2 \over \pi \rho_0 } =  2 B_{11} 
+ { 2 M_c  \over a_1} C \left( e 
\right ) \, ,
\ee
where 
\be
\label{141}
C (e) = P ( e) - { 1 \over 9} E(e) \, , 
\ee
and the leading term $2 B_{11}$ formally coincides with the Newtonian 
expression (\ref{030}). 
The equation for $C(e)$ has been derived by using MAPLE, using again the 
derivatives given in LRS [expressions (A3), (A9)] and 
the equations reported in Appendix D. The value of the 
eccentricity at the secular 
instability point, $e_{sec}$, is calculated by equating the two expressions 
(\ref{122}) and (\ref{140}) and is reported in Table \ref{tab3}. Figure 
\ref{fig6} shows the ratio $T/ |W|$ evaluated at $e=e_{sec}$, 
as a function of the 
compactness parameter $M/R_s$. Squares in Figure \ref{fig3} mark the 
instability as a function of the two gauge invariant ratios 
$\Omega^2/(\pi \rho_0)$ and $T/ |W|$, and thus separate the regions of 
stable and unstable configurations. As we can see, 
effect of general relativity is to move the instability point to an 
eccentricity larger than what Newtonian theory predicts. This 
corresponds also to larger values of the two ratios $\Omega^2 / (\pi 
\rho_0)$ and $T / |W|$ (see Figure \ref{fig3}), showing, therefore, that 
relativistic gravitation tends to stabilize a star against secular 
instability to a Jacobi-like, nonaxisymmetric bar mode. 
\placetable{tab3}
\placefigure{fig6}

\section{Discussion and Conclusions}  

In PN gravitation the rotational velocity along the 
equilibrium sequence is not much different from that one derived in the 
Newtonian 
limit, although 
a star of a given eccentricity has a slightly larger value of $\Omega^2/(\pi 
\rho_0)$. 
General relativity, however, is more crucial in influencing the stability 
properties. According to our treatment, the 
critical value of the eccentricity for the onset of the 
bar instability increases as the star becomes more 
relativistic, in the regime in which the PN approximation is valid. Both 
invariant 
ratios, $\Omega^2/(\pi \rho_0)$ and $T/|W|$, also increase at the onset 
of instability above the values found in Newtonian theory. Gravitational 
radiation does not enter at PN order; it is present only at the 2.5 PN 
level and higher. The secular instability we have identified 
is driven by the presence of viscosity. That conclusion 
is consistent with our variational procedure, whereby the rotation was 
kept uniform and angular momentum was conserved. Viscous dissipation 
conserves angular momentum and drives a star to uniform rotation. 
Gravitational radiation dissipation conserves circulation (Miller 1974; 
LRS), not angular momentum, and does not maintain uniform rotation. 
The presence of a stabilizing effect due to general relativity on the 
Jacobi--like bar 
mode instability is in agreement with the BFG and 
Bonazzola, Frieben \& Gourgoulhon (1997) analysis of relativistic 
polytropes, who observe a growth of the critical adiabatic index 
when the relativistic character of the configuration is increased. 
Shear viscosity can provide a such dissipation mechanism in a very cold NS 
($T \lesssim 10^6$ K), but it is inefficient for hot, newborn objects 
with $T\sim 10^{10}$ K (see e.g. BFG), 
where the nonaxisymmetric instability is more probably 
induced by gravitational radiation (the CFS instability) and proceeds via a 
Dedekind--like mode. In this case, the fully relativistic SF computations 
show that effects of general relativity are reversed, and that the 
instability is significantly strengthed with respect to Newtonian theory. 
SF calculations are fully relativistic, so this 
bimodal behaviour may be explained in terms of strong field effects. 
However, as originally suggested by SF, there is also the likelihood 
that, in general relativity, the viscosity driven and the 
gravitational radiation driven $m=2$ modes may no longer coincide and 
that the gravitational field plays a different role in each of them. 
Although at present no firm conclusion can be reached, this speculation 
is further strengthed by the results of our PN treatment. 
Unfortunately, a direct comparison between our results and the ones 
obtained by BFG and SF is not possible, since in both that cases 
numerical constraints prevented them from treating incompressible fluid 
configurations, such as the Maclaurin and Jacobi ellipsoids, where the 
density profile is strongly discontinuous at the surface. In the BFG's 
case, numerical tests showed that the density steepens dramatically 
(and the numerical error increases more and more) for $\gamma > 3.25$ ($n 
< 0.44$). Models presented by SF are restricted to values of the polytropic 
index $n \geq 1$, since the presence of 
discontinuous derivatives of energy density 
and metric functions across the surface of the configuration make the 
description less accurate for stiffer equations of state. Moreover, the 
$m=2$ mode is only present in the SF model with $n = 1$. 

In Newtonian theory, the variational principle also has been used by LRS to 
investigate the 
secular instability to dissipation which conserves 
circulation $\cal C$, rather than $J$, such as the emission of 
gravitational waves. In this case they used 
the energy functional for Dedekind ellipsoid, rather than a Jacobi 
ellipsoid as in the case of a viscosity--driven mode. Because the energy 
function of a Riemann-S ellipsoid is symmetric under interchange of $\cal 
C$ and $J$, the two analysis are virtually identical in Newtonian theory, 
with $\cal C$ appearing in place of $J$ in all results. The 
previous discussion suggest that this symmetry may be broken in a 
relativistic treatment; this may be apparent at PN order and 
we plan to investigate this issue in a forthcoming work. 

\acknowledgments

We are grateful to J.L.~Friedman and N.~Stergioulas for providing a copy 
of Stergioulas's Ph.D Thesis prior to publication. We thank also Thomas 
Baumgarte for several very helpful discussions. This work was supported 
by NSF Grants AST 96--18524 and NASA Grant NAG 5--3420 at the University 
of Illinois at Urbana--Champaign.

\appendix 

\section{Comparison with the Chandrasekhar's Results}

Results presented here have been derived by using a 3+1 
decomposition of the Einstein field equations. It is therefore useful to 
compare the resulting expressions with those obtained by CN, who worked 
with the usual field equations. In 
the first PN approximation, the metric coefficients $g_{00}$, 
$g_{0 \alpha}$ and $g_{\alpha \beta}$ are retained up to the 
order $O(c^{-4})$, $O(c^{-3})$ and $O(c^{-2})$, respectively (see CN 
and footnote \ref{fo1} in this paper). 
In this Appendix we report the comparison. In section A.1 we 
show that the same gauge condition is common to both treatments. In section 
A.2 we compare the metric functions, while in A.3 we compare the 
conserved quantities. 

\subsection{{\it The Spatial Gauge Condition}}

Here we verify that the gauge condition adopted in CN is identical to 
our own. Chandrasekhar's expressions are obtained in the gauge 
\be 
\label{58}
\sum_i \der{P_i} {x_{i} } = - 3 \der {U}{t} + O \left ( 1 \over c^{-2} 
\right ) \, , 
\ee
where $P_{i} = - \beta^i_{PN}$ [see CN, equations (3) and (6); section 
A.2.2 in this paper]. In our approach the condition is [see CST96, equation 
(3)] \be \label{59}
\der{ \ln
\gamma^{1/2}}{t} = 
D_k \beta^k \, , 
\ee    
where $\gamma = \Psi^{12}$. At leading order, we have the approximate 
expressions
\be
\label{60}
D_k \beta^k  \approx \sum_k \der{ \beta^{k}_{PN}}{x_k} \, , 
\quad  \der{ \ln
\gamma^{1/2}}{t} \approx 
{6 \over \gamma} \der{\Psi}{t} \approx 
6  \der{\Psi}{t} \, . 
\ee
Substituting $\Psi = 1 - \Phi_{N}/2$ and neglecting terms of higher 
order, we obtain 
\be \label{61} 
-3 \der
{\Phi_N}{t} \approx \sum_k \der{\beta^{k}_{PN}}{x_k} 
\, , 
\ee
which coincides with the Chandrasekhar's choice (\ref{58}). 

\subsection{{\it The Metric}}

\subsubsection{{\it The Lapse Function and the Conformal Factor}}

We can now compare expressions (\ref{31}) and (\ref{32}) for 
$\alpha_{N}$ and $\alpha_{PN}$ with the corresponding terms derived by 
CN. We start from the CN's expression of $g_{00}$  
\be
g_{00} \approx 1 - 2 U + 2 \left ( U^2 - 2 \Phi_{Ch} \right ) \, , 
\ee
where $\Phi_{Ch}$ is the solution of
\be
\nabla^2 \Phi_{Ch} = - 4 \pi \rho_0 \left ( v^2 + U + {3 \over 2} {P 
\over \rho_0 } \right ) 
\ee
and $U = - \Phi_N$ [see CN, equations (3) and (4)]. 
Note that CN used a different signature 
for the metric, i.e. + - - -. At our order of approximation, we find 
\bea   
\label{alpha}
\alpha_{Ch} &\approx & \sqrt{g_{00}} \approx 
1 - U + U^2 - 2 \Phi_{Ch} - {U^2 
\over 2} \\ \nonumber  
& \approx & 
1 - U + {U^2 \over 2} - 2 \Phi_{Ch} \, . 
\eea
On the other hand, $\Phi_{PN}$ and $\Theta_{PN}$ are defined as the 
solutions of [equation (\ref{22}), (\ref{28})]
\be
\label{22app}
\nabla^2 \Phi_{PN} = - 10 \pi \rho_0 \Phi_N  + 4 \pi \rho_0 v^2
\, , 
\ee
\be 
\label{28app}
\nabla^2 \Theta_{PN} = 6 \pi \rho_0 \left ( v^2 + 2 
 { P \over \rho_0 } - { 1 \over 2} \Phi_N \right ) \, , 
\ee
so that, by comparing the corresponding differential equations, we can 
immediately recognize that $-\Phi_{Ch} = \Phi_{PN}/4 + \Theta_{PN}/2$. 
Substituting this result into (\ref{alpha}) gives 
\be
\alpha_{Ch} \approx 1 + \Phi_N + {\Phi_N^2 \over 2} + {\Phi_{PN} \over 2} 
+ \Theta_{PN} \, , 
\ee
which agrees with the solution derived in section 4.1. 

For the spatial metric the comparison is trivial. The PN expression in CN 
is [CN, equation (3)] 
\be 
\label{gab}
g_{ab} \approx - \left ( 1 + 2U \right ) \delta_{ab}\,  , 
\ee
and, and the PN order, $\Psi^4 = - g_{ab} \approx \left (1 + U/2 \right ) 
\delta_{ab}$. 

\subsubsection{{\it The Shift Function}}

Consider now the shift vector. In the CN's formalism, at the PN order it is 
[CN, equations (3), (5)] 
\be
\label{s1}
g_{0a} \approx g^{0a} \approx P_a \, , 
\ee
where $P_a$ is the solution of the equation 
\be \label{s2}
\nabla^2 P_a = - 16 \pi \rho_0 v_a + \ders{U}{x_a}{t} \, . 
\ee
Using the gauge condition [CN, equation (6)] 
\be 
\label{s3}
\der{U}{t} \approx - {1 \over 3} 
\sum_i \der{P_i} {x_{i} } \, , 
\ee
and neglecting terms of higher order, expression (\ref{s2}) can be 
rewritten as 
\be \label{s4}
\nabla^2 P_a \approx   - 16 \pi \rho_0 v_a 
- {1 \over 3}
 \der{\, }{x_a}
\sum_i \der{P_i} {x_{i}}
\, . 
\ee
On the other hand, in our formalism we have [equations (\ref{7}), (\ref{9}), 
(\ref{37}), (\ref{38})] 
\bea
\label{s5}
\nabla^2 \beta^i & = & \nabla^2 G^i - { 1 \over 4 } \nabla^2 \left 
(\nabla^i B \right ) \\ \nonumber 
& \approx & 16 \pi \rho_0 v^i - { 1 \over 4 } \nabla^i \nabla_k G^k 
\, . 
\eea
Taking the divergence of $\beta^k$ and using equation (\ref{9}) again gives 
\be 
\label{s6}
\nabla_k \beta ^k = \nabla_k G^k - { 1 \over 4} \nabla^2 B = 
{ 3 \over 4} \nabla_k G^k \, , 
\ee
which can be substituted into (\ref{s5}), yielding $P_{i} = - 
\beta^i_{PN}$. 

\subsection{{\it The Conserved Quantities}}

The derivation of 
conservation laws in general relativity 
has been considered by 
Chandrasekhar (1969b) and CN, who 
used the symmetric energy--momentum complex of Landau--Lisfhitz to 
determine integral forms of conserved quantities at 
various post--Newtonian orders.\footnote{Actually, in the first 
PN approximation the same expressions were originally obtained by 
Ch65a from a direct inspection of the PN generalization of the equations of 
motion and assuming the supplementary condition for isentropic flow. 
However, the fact that it is possible to derive the conserved 
quantities without the aid of the Landau--Lifshitz complex (at least when 
the additional condition is provided) is a peculiarity of the first PN 
order. The method discussed in Chandrasekhar (1969b) appears to be the 
most convenient way, allowing the derivation of the integral quantities 
in all the PN approximations beyond the first.} 
In this Appendix we compare our results (\ref{mmo}) and (\ref{87}) with the 
corresponding integrands obtained by CN. 
Consider the conserved energy. In the first PN approximation, 
CN derive the energy per unit volume of the fluid  [see CN, equation 
(67)] 
\bea
\label{c}
{\cal E} &= &\rho_0 \left( { 1 \over 2} v^2 - { 1\over 2} U + \Pi \right 
) \\ \nonumber 
&+& \rho_0 \left [
{ 5 \over 8 } v^4  
+ { 5 \over 2 } v^2 U  
- { 5 \over 2 } U^2 + 2 U \Pi + v^2   \left ( \Pi + { P \over \rho_0 } 
\right ) - 
{ 1 \over 2} v_i P_i \right ] 
\, , 
\eea
where $\Pi$ is the internal energy and $P_i = - \left (\beta_i 
\right)_{PN}$. In the incompressible case of interest here, $\Pi = 0$. 
In order to compare (\ref{c}) with the integrand function appearing in 
our expression (\ref{mmo}), we note that CN's results have 
been derived exploiting a set of identities between quantities that are 
equal, modulo a divergence. In fact, upon integration, two functions that 
are equal modulo divergence give the same conserved quantities (see 
Chandrasekhar 1969b for a detailed discussion). 
After some lengthy reductions, it is 
possible to demonstrate that the $K^{ij}K_{ij}$ contribution appearing   
in (\ref{mmo}) can be rewritten as 
\be 
\label{88} 
K^{ij}K_{ij} = { 1\over 2} \left ( 
\der{P_a}{x_b}
\der{P_a}{x_b}
+
\der{P_a}{x_b}
\der{P_b}{x_a } \right ) - { 1 \over 3} \left ( {\rm div} \vec P \right)^2 
\, . \ee
From equations (62), (63) of CN we have  
\be 
\label{89} 
\der{P_a}{x_b}
\der{P_a}{x_b} \equiv 16 \pi \rho_0 v_i P_i - 3 \left ( \der {U }{t} \right 
)^2 \quad {\rm ( mod ~ div )} \, , 
\ee
\be 
\label{90} 
\der{P_a}{x_b}
\der{P_b}{x_a} \equiv 9 \left ( \der {U }{t} \right )^2 
\quad {\rm ( mod ~ div )}\, ,  
\ee
and these equalities can be used, together with the gauge 
condition (\ref{58}) 
\be
\label {91}  
- 3 \der {U}{t} = {\rm div} \vec P \, , 
\ee
to rearrange expression (\ref{88}). This yields
\bea
\label{92} 
K^{ij}K_{ij} 
&\equiv & 8 \pi \rho_0 v_i P^i 
+ { 1 \over 3} \left ( {\rm div} \vec P \right )^2
- { 1 \over 3} \left ({\rm div} \vec P \right )^2
\quad {\rm ( mod ~ div )} \\ \nonumber 
&\equiv & - 8 \pi \rho_0 \left ( v_i \beta^i \right )_{PN}
\quad {\rm ( mod ~ div )} \, . 
\eea
Finally, inserting back (\ref{92}) into (\ref{mmo}) one 
recognizes that our expression for the energy 
density agrees with equation (\ref{c}).   

Consider next the angular momentum. Start from expression 
(141) by Ch65a for the conservation of total angular momentum,  
\be 
\label{93} 
\int_V \left ( 
x_i \pi_j - x_j \pi_i \right ) d^3 x  
= constant  \, , 
\ee 
where 
\be \label {94} 
\pi_i = \sigma v_i + { 1 \over 2} \rho_0 \left ( U_i - U_{k;ik} \right ) 
+ 4 \rho_0 \left ( v_i U - U_i \right ) \, ,  
\ee
$U_i$, $U_{k ; ik }$ are quantities which enter the shift function, 
$\sigma = \rho_0 \left (1 + v^2 + 2 U + P / \rho_0 \right )$, and where 
incompressibility implies $\Pi = 0$. Specializing to our velocity profile, 
we have $\pi_3 = 0$ and 
\be \label{jch} 
J_{Ch} = \int_V \left ( 
x_1 \pi_2 - x_2 \pi_1\right ) d^3 x  \, . 
\ee
After some algebra it is possible to demonstrate that the PN expression of 
the shift function in Ch65a is 
\be 
\label {95} 
\left ( \beta_{PN}^i \right )_{CH}
\approx - \alpha^2 g^{0 i } 
\approx - g^{0 i } = - 4 U_i + { 1 \over 2} \left ( U_i - 
U_{k;ik} \right ) = - { 7 \over 2} U_i - { 1 \over 2}U_ { k;ik } \, ,  
\ee
and that the integrand appearing into (\ref{jch}) can be rearranged as 
\bea 
\label{96} 
x_1 \pi_2 - x_2 \pi_1  
&= & { \rho_0 \over \Omega } \left ( v^2 + v^4 + 6 U 
v^2 + { P \over \rho_0 } v^2 \right ) \\ \nonumber 
& - & { 7 \over 2 } \rho_0 x_1 U_2 - { 1\over 2} \rho_0 x_1 U_{k;2k} 
+ { 7 \over 2 } \rho_0 x_2 U_1 + { 1\over 2} \rho_0 x_2 U_{k;1k} + \dots 
\, . \eea 
Rewriting the last row in the previous formula as  
\be 
\label{97} 
\rho_0 x_1 \beta^2 - \rho_0 x_2 \beta_1 = 
{\rho_0 \over \Omega}  \left ( v_2 \beta^2 + v_1 \beta^1 \right ) 
\ee 
and inserting (\ref{96}), (\ref{97}) into the 
conserved quantity (\ref{jch}), we finally recover the same 
expression as in (\ref{87}). 

\section{The Spherical Limit}

\subsection{{\it Comparison with the Exact Solution}}

In the spherical limit, 
the interior metric for a homogeneous 
configuration admits an exact solution in full general 
relativity. In this Appendix we 
verify that 
our results agree with the PN expansion of the exact solution.
We start from the interior 3--metric, written in Schwarzschild and 
in conformal (isotropic) coordinates as
\be
\label{b1}
d s^2_{\left( 3\right )} =  
{ 1 \over  1 - 2 m \left ( r_s \right ) / r_s } d r_s^2 + 
r_s^2 d \Omega^2 
=  \Psi^4 \left [ d r^2 + r^2 d \Omega^2 \right ] \, ,  
\ee
where $m (r_s) = 4 \pi \rho_0 r^3_s/3$. Equating the metric coefficients 
gives \bea 
\label{b2}
\Psi^2 dr &=&  { 1 \over  \sqrt { 1 - 2 m \left ( r_s \right ) / r_s } } 
dr_s \\  
\label{b3}
r_s & = & \Psi^2 r \, , 
\eea 
and these relations can be combined to yield the differential equation 
\be 
\label{b4}
{ 2 \over \Psi} d \Psi = \left ( 1 - 
{ 1  \over  \sqrt { 1 - 2 m \left ( r_s \right 
) / r_s } } \right ) { dr_s \over r_s } \, . 
\ee
Integrating (\ref{b4}) gives
\be 
\label{b5}
\Psi = K \left ( 1 + \sqrt { 1 - { 8 \pi \rho_0 \over 3} r_s^2 } \right 
)^{1/2} \, , 
\ee
where $K$ is a constant of integration. By using (\ref{b3}), expression 
(\ref{b5}) can be written in terms of isotropic coordinates as 
\be 
\label{b6}
\Psi = { \displaystyle{ \sqrt2 K} \over \displaystyle \left ( 1 + 
K^4 { 8 \pi \rho_0 \over 3} r^2 \right )^{1/2} } \, . 
\ee
The value of $K$ is determined by 
matching the solution with the exterior metric. This is [see Misner, 
Thorne \& Wheeler, equation (31.22), page 840]
\be
\label{b7}
\Psi = 1 + { M \over 2 r} \, , 
\ee 
and, at the surface of the star, $r = R$. Imposing the 
boundary condition yields 
\be 
\label{b8}
\sqrt{2} K =  \left ( 1 + {M \over 2R} \right )^{3/2}    
\, . \ee
The PN expression of $\Psi$ can be then derived expanding (\ref{b6}), and 
using the relations 
\be 
\label{b9}
M = { 4 \pi \rho_0 \over 3 } R_s^3 \, , 
\ee
\be 
\label{b10}
R_s = R \left ( 1 + { M \over 2 R } \right )^2\, . 
\ee
This yields 
\be 
\label{b11}
\Psi \approx 1 + \pi \rho_0 R^2 - { \pi \rho_0 \over 3} r^2 + 
{ 25 \over 6} \left (  \pi \rho_0 \right )^2 R^4 - 
{ 5 \over 3 } \left (  \pi \rho_0 \right )^2 R^2 r^2 + 
{ 1 \over 6 } \left (  \pi \rho_0 \right )^2 r^4 \, . 
\ee
By specializing our solution (\ref{41}) and (\ref{46}) to the spherical case 
gives 
\be 
\label{b12}
\Phi = \Phi_N + \Phi_{PN} = - 2\pi \rho_0 R^2 +  { 2 \pi \rho_0 \over 3} 
r^2 - { 25 \over 3} \left (  \pi \rho_0 \right )^2 R^4 +
{ 10 \over 3 } \left (  \pi \rho_0 \right )^2 R^2 r^2 -
{ 1 \over 3 } \left (  \pi \rho_0 \right )^2 r^4 \, .
\ee
which, with (59), agrees with (\ref{b11}). 

\subsection{{\it The Relativistic Correction to the Total Energy}}

Our integrated expression for $M-M_0$ allows a comparison with the 
relativistic correction to the total energy of a spherical configuration 
as derived by ST (see also 
Zel'dovich \& Novikov, 1971). Following ST, let us define the 
relativistic correction as 
\be 
\label{105} 
\Delta E_{GTR}  \equiv M - M_0 - E_{N} \, , 
\ee 
where $E_{N}$ is the Newtonian gravitational energy
\be 
\label{105a} 
E_{N}  = 
- { 3 \over 5 } { M_N^2 \over R_N } = 
- { 3 \over 5 } 
\left (4 \pi \over 3 \right )^2 \rho_0^2 R_N^5
\, , 
\ee 
and $M_N$, $R_N$ are the Newtonian mass and radius, respectively.
Dimensionally, at PN order this correction can be written as 
\be  
\label{106} 
\Delta E_{GTR} = - k  M^{7/3} \rho_0^{2/3} \, , 
\ee 
and the value of $k$ in the homogeneous case can be obtained by 
using equation (6.9.29) in ST for an incompressible gas with polytropic 
index $n=0$. This yields\footnote{Note that, in 
the limit $n\to 0$, the two integrals appearing in ST (6.9.29) become 
$I_1 = - 3 M^3 / (14 R_N^2)$ and $I_2 = - 3 M^3 / (35 R_N^2)$. 
This gives 
$$
\Delta E_{GTR} = - { 3 \over 70 }{ M^3 \over R_N^2} =    - {3 \over 70} 
\left (4 \pi \over 3 \right )^{2/3} M^{7/3} \rho_0^{2/3} \, .
$$} 
\be
\label{107} 
k = {3 \over 70} 
\left (4 \pi \over 3 \right )^{2/3} \, . 
\ee
In order to recover this result from our equations, we specialize 
equation (\ref{103d}) to the nonrotating, 
spherical case. In this limit $A_1=A_2=A_3=2/3$ and $f=1$, so that 
\bea
\label{108} 
M - M_0 & \approx & 
- { 3 \over 5} {M_c^2 \over R} - { 51 \over 14} {M_c^3 \over 
R^2}\\ \nonumber & = & 
- { 3 \over 5} 
\left ( { 4 \pi \over 3 } \right )^2 \rho_0^2 R^5 
- { 51 \over 14} 
\left ( { 4 \pi \over 3 } \right )^3 \rho_0^3 R^7
 \, . 
\eea
On the other hand, the exact relation between the conformal (isotropic) 
radial coordinate $R$ and the Schwarzschild radius $R_s$ is given by 
(\ref{b10}). At PN order 
\be
\label{109a} 
R \approx  R_s \left ( 1 - {M \over R_s }\right ) \, ,  
\ee
and we left to derive the relation between $R_s$ and $R_N$. 
Start from the definition (6.9.12) in ST  
\be 
\label{rnv}
 R_N \equiv \left ( {3 {\it V} \over 4 \pi } \right )^{1/3} \, ,  
\ee
where ${\it V}$ is the proper volume. 
To first order [see (6.9.15)] 
\bea 
\label{rn}
R_N &\approx  &
R_s \left ( 1 + { 1 \over R_s^3 } \int_0^{R_s} m r_s d r_s \right ) \\ 
\nonumber
 &\approx  & R_s \left ( 1 + { 1 \over R_s} 
{ 4 \pi \rho_0 \over 3} \int_0^{R_s} r_s^4 dr_s \right ) \approx R_s \left 
( 1 + { M \over 5 R_s } \right ) \, . 
\eea
This gives 
\be
\label{109b} 
R_s \approx R_N \left ( 1 - {M \over 5 R_N }\right ) \, ,  
\ee
and, combining (\ref{109a}) and (\ref{109b}), yields 
\be
\label{110} 
R  \approx R_N \left ( 1 - {6 \over 5} {M \over R_N }\right )  \, . 
\ee
Finally, substituting the equation (\ref{110}) into equation (\ref{108}) 
and subtracting $E_{N}$, we obtain the relativistic correction 
\be 
\label{112} 
M - M_0 - E_{N} \approx 
- { 3 \over 70 }{ M_N^3 \over R_N^2} \approx 
- { 3 \over 70 }{ M^3 \over 
R_N^2} \, , \ee
which agrees with the ST's result. 

\section{The Integral ${\cal I}_{10}$}

The integral ${\cal I}_{10}$ has been obtained by 
using MAPLE. We only report here the resulting expression, that is  
\bea
\label{i10}
\tilde {\cal I}_{10} & = & 
(39/560 ) (A_1-A_2)^2/\left (\lambda_1 \lambda_2 \right ) 
-(9/140 )\overline A_{12}^2(1-\lambda_1^3/\lambda_2^3)(
\overline A_{122}-\overline A_{112}\lambda_1^3/\lambda_2^3)\lambda_1^2/
\lambda_2^4 \\ \nonumber
& + & (3/2240)(A_1-A_2)^2(
9\overline A_{112}^2 +\overline 
A_{122}^2 +(-2A_2\lambda_2^3+2\lambda_2^3-A_1\lambda_2^3 \\ \nonumber 
&-&\overline A_{12}\lambda_1^3+\overline 
A_{12}\lambda_1^3\lambda_2^3)^2/(1- \lambda_2^3)^2/\lambda_1^6/
(1- \lambda_1^3)^
2)\lambda_1^5/\lambda_2^7
\\ 
\nonumber 
&+& (3/8) 
(\lambda_1^2/\lambda_2^4)
(-1/5(A_1-A_2\lambda_1^3/\lambda_2^3)^2
 + (3/35)\lambda_1 ^6(-1/(1-\lambda_1^3)(2A_1-2+A_2)\\ 
\nonumber &+& 1/(1-\lambda_2^3)(2A_2-2+A_1))
^2+(6/35)\lambda_1^3(-1/(1-\lambda_1^3)(2A_1-2+A_2)+1/(1-\lambda_2^3)
\\ \nonumber 
&\times & (2A_2 -2 +
A_1))(A_1-(1/3)\overline A_{12}(\lambda_1^3/
\lambda_2^3)+1/3/(1-\lambda_1^3)(2A_1-2+A_2 ) \lambda_1^3
\\ \nonumber 
&-&(\lambda_1^3/\lambda_2^3)(A_2 -
(1/3)\overline A_{12}+1/3/(1-\lambda_2^3)(2A_2-2+
A_1)\lambda_2^3)))+(3/280)\overline A_{12}(1 \\ 
\nonumber &-&\lambda_1^3/\lambda_2^3)(\lambda_1^2/\lambda_2^4)(3\overline 
A_{12}(A_1-A_2\lambda_1^3/\lambda_2^3)-9
A_1\overline A_{112}\\ 
\nonumber &+&9A_2 (\lambda_1^6/\lambda_2^6)
\overline A_{122}+3(\overline A_{112}-
\overline A_{122}\lambda_1^3/\lambda_2^3)\overline 
A_{12}\lambda_1^3/\lambda_2^3
\\ \nonumber &-&\lambda_1^3
/(1-\lambda_1^3)(2A_1-2+A_2)(3\overline 
A_{112}-(-2A_2\lambda_2^3+2\lambda_2^3-
A_1\lambda_2^3-\overline A_{12}\lambda_1^3 \\ \nonumber 
& +& \overline 
A_{12}\lambda_1^3 \lambda_2^3)/(-1+
\lambda_2^3)/(-1+\lambda_1^3))+\lambda_1^
3/(1-\lambda_2^3)(2A_2-2+A_1)(3 \overline A_{122} \lambda_1^3/\lambda_2^3
\\ \nonumber 
&-&(-2 A_2 \lambda_2^3 
+2 \lambda_2^3-A_1 \lambda_2^3-\overline A_{12} \lambda_1^3+\overline 
A_{12} \lambda_1^3 \lambda_2^3)/(-1+\lambda_2^3)/(-1+
\lambda_1^3)))\\ 
\nonumber &+& (3/70)(\lambda_2^2/\lambda_1) (1/(1
-\lambda_2^3)^2 (2A_2-2+A_1)^2 \lambda_1^3/\lambda_2^3+1/(1-
\lambda_1^3)^2 (2 A_1-2+A_2)^2) \\ \nonumber &+& (3/1120) \overline 
A_{12}^2 (1-\lambda_1^3/\lambda_2^3)^2 (1+\lambda_1
^3/\lambda_2^3) (-2 A_2 \lambda_2^3+2 \lambda_2^3-A_1 \lambda_2^3
-\overline A_{12} \lambda_1^3 \\ 
\nonumber & +& \overline A_{12} 
\lambda_1^3 \lambda_2^3)^2/(-1+\lambda_2^3)^2/\lambda_1/(-1
+\lambda_1^3)^2/\lambda_2^4-(3/140 )\overline A_{12} (1 
\\ \nonumber & -& \lambda_1^3/\lambda_2^3) (-
1/(1-\lambda_1^3) (2A_1-2+A_2)+\lambda_1^3/\lambda_2^3/(1
-\lambda_2^3) (2 A_2-2+
A_1
)) (-2 A_2*\lambda_2^3 \\ \nonumber & +& 2 \lambda_2^3-A_1 \lambda_2^3
-\overline 
A_{12} \lambda_1^3+\overline A_{12} \lambda_1^3 
\lambda_2^3)/(-1+\lambda_2^3)/\lambda_1/(-1+\lambda_1^3)/\lambda_2 
\\ \nonumber & +& (3/128 ) \overline 
A_{12}^2 (1-\lambda_1^3/\lambda_2^3)^2 (\lambda_1^2/\lambda_2^4)(
-\overline A_{12}^2+(6/7 )\overline A_{12} \overline A_{112}
+(9/35 )\overline A_{112}^2 \\ \nonumber & +& (2/7 )\overline A_{12}
(-2 A_2 \lambda_2^3+2 \lambda_2^3-A_1 \lambda_2^3-\overline A_{12}
 \lambda_1^3 \\ \nonumber & 
+& \overline A_{12} \lambda_1^3 \lambda_2^
3)/(-1+\lambda_2^3)/(-1+\lambda_1^3)+(1/35)(-2 A_2 \lambda_2^3+2 
\lambda_2^3-A_1 \lambda_2^3
-\overline A_{12} \lambda_1^3 \\ \nonumber & +& \overline 
A_{12} \lambda_1^3 \lambda_2^3)^2/(-1+\lambda_2^3)^2/(-1+\lambda_1^3)^2 
+ (27/35 )\overline A_{122}^2 \lambda_1
^6/\lambda_2^6 
 - (6/35) \overline 
A_{122} (-2 A_2 \lambda_2^3 \\ \nonumber & +& 2 \lambda_2^3-A_1 \lambda_2^3  
- \overline A_{12} 
\lambda_1^3+\overline A_{12} \lambda_1^3 \lambda_2^3)/(-1
+\lambda_2^3) \lambda_1^3/(-1 
\\ \nonumber & +& \lambda_1^3)/\lambda_2^3)+ (3/10) \overline A_{12}^2
\lambda_1^2/\lambda_2^4+(81/280 ) 
(\lambda_2^2/\lambda_1^4 ) (A_1-(1/3 )\overline 
A_{12} \lambda_1^3/\lambda_2^3+1/3/(1-\lambda_1^3) (2
 A_1 \\ \nonumber & -& 
2+A_2) \lambda_1^3)^2+ (81/280 ) (\lambda_1^2/\lambda_2^4 
)(A_2- (1/3 ) \overline A_{12} 
+1/3 /(1-\lambda_2^3) (2 A_2-2+A_1) \lambda_2^3)^2
\\ \nonumber 
&-& (27/140 ) (A_1 
- (1/3 ) \overline A_{12} 
\lambda_1^3/\lambda_2^3+1/3/(1-\lambda_1^3)(2A_1-2+A_2) \lambda_1^3) (A_2
-(1/3 )\overline 
A_{12}+1/(3
\\ \nonumber & -& 3 \lambda_2^3) (2 A_2 
- 2+A_1) \lambda_2^3)/\lambda_1/\lambda_2- (9/70 ) \overline A_{12} (
A_1
+A_2 \lambda_1^3/\lambda_2^3)/\lambda_1/\lambda_2+ (3/70 ) \overline 
A_{12} \lambda_1^2
\\ 
\nonumber 
& \times &(-1/(1 
-\lambda_1^3) (2 A_1- 2+A_2)-1/(1-\lambda_2^3) (2 A_2-2+A_1))/\lambda_2, 
\eea
where 
\be
\label{ove}
\nonumber 
\overline A_{12} = a_1^2 A_{12} 
\, , \quad \quad 
\overline A_{122} = a_1^4 A_{122} 
\, , \quad \quad 
\overline A_{112} = a_1^4 A_{112} \, ,  
\ee
only depend on the axial ratios. 

\section{The Index Symbols} 

The expressions of the metric functions we have derived have been 
presented in terms of the index symbols $A_{ijk \dots}$ and $B_{ijk 
\dots}$, where  
\be 
\label{defis}
A_{ijk \dots}
= a_1 a_2 a_3 \int_0^\infty {du \over \Delta
(a_i^2 + u ) 
(a_j^2 + u ) 
(a_k^2 + u ) \dots }
\, , \ee
\be 
B_{ijk \dots}
= a_1 a_2 a_3 \int_0^\infty {u du \over \Delta
(a_i^2 + u ) 
(a_j^2 + u ) 
(a_k^2 + u ) \dots }
\, ,  \ee
and where $ \Delta^2 = (a_1^2 + u ) (a_2^2 + u ) (a_3^2 + u )$. 
The use of these quantities is particularly convenient, since it allows 
for a more compact form of the results. 
In the expressions of the conserved quantities, however, it is more 
convenient to 
reduce the number of different index symbols entering in the final forms. 
This is because when we perform the variations of the conserved quantities 
by using MAPLE, the derivatives of the index symbols are evaluated 
analytically and then substituted in the variations. 
The index symbols are 
not independent: they are manifestly symmetric in their indices and obey 
to a set of identities and recursive relations (see Chapter 3 in Ch69). 
Exploiting these properties we derived the results presented in D.1; 
these have been used to obtain the final expressions of the $g_i$ and 
$p_i$ functions which enter in $M$, $M_0$ and $J$ [see 
(\ref{104})--(\ref{104app}), (\ref{i10})]. Note that the final forms only 
depends on the quantities 
\be 
A_i , \overline A_{12}, \overline A_{122}, \overline A_{112}\, . 
\ee
Then, variations have been performed by using a MAPLE program. 
The final results have been specialized to axisymmetry, 
using expressions (\ref{oveas1}), (\ref{oveas2}) and substituting the 
derivatives given in D.2. 

\subsection{{\it The Coefficients $A_{ijk}$ and $B_{ijk}$; General 
Expressions}}

The $A_i$'s functions are given in terms of the standard 
(elliptic) integrals by expressions (3.33)--(3.35) in Ch69. The other index 
symbols can be written as   
\bea
A_{il} & = & -
{ 1 \over a_i^2 - a_l^2 } \left ( A_i - A_l 
\right) 
\hspace {6. cm}  {\rm for}~i\neq l 
\\ \nonumber & \,& 
\\ \nonumber 
A_{ii} & = &
{ 2 \over 3 a_i^2 }  - 
{ 1 \over 3 } A_{il} -  
{ 1 \over 3 } A_{im} 
\hspace {6. cm}  
 {\rm for}~i\neq l \neq m \\ 
& = & { 2 \over 3 a_i^2 }  - 
{ 1 \over 3 } \sum_j \left ( 1 - \delta_{ij} \right ) 
A_{ij}
\\ 
\nonumber & \,& 
\\ 
A_{ilm} & = &-{ 1 \over a_i^2 - a_l^2 } \left ( A_{im} - A_{lm} 
\right) 
\hspace {5.5 cm}  
{\rm for}~i\neq l \\ 
\nonumber & \,& 
\\ 
A_{ill} & = & -\displaystyle{ 1 \over a_i^2 - a_l^2 } \left ( A_{il} - 
A_{ll} \right)  
\hspace {5.8 cm}  
 {\rm for}~i\neq l \\ 
\nonumber & \,& 
\\ \nonumber 
A_{iil} & =  & 
\displaystyle{ 5\over 3 a_i^2 } A_{il} - 
\displaystyle{ a_l^2\over a_i^2 } A_{ill} - {A_{ilm} \over 3 }  
{ a_m^2 \over a_i^2} 
\hspace {4.8 cm}  
{\rm for}~i\neq l \neq m \\ 
& =  & 
\displaystyle{ 5\over 3 a_i^2 } A_{il} -  
\displaystyle{ 1\over 3  a_i^2 } \sum_m A_{ilm} a_m^2 \left ( 1 - 
\delta_{mi} \right ) \left ( 1 + 2 \delta_{lm} \right )  
\hspace {2. cm}  {\rm 
for}~i\neq l
\\ \nonumber & \,& \\ \nonumber 
A_{iii} 
& =& \displaystyle{ 1 \over a_i^2} A_{ii} 
-\displaystyle{ a_l^2 \over 5 a_i^2} A_{iil} 
-\displaystyle{ a_m^2 \over 5 a_i^2} A_{iim} 
\hspace {4.8 cm} 
{\rm for}~i\neq l \neq m \\ 
& =& \displaystyle{ 1 \over a_i^2} A_{ii} 
-\displaystyle{ 1 \over 5 a_i^2} \sum_m A_{iim} a_m^2 \left ( 1 - 
\delta_{im} \right ) \, . 
\eea
The  $B_{ilm..}$'s are then: 
\be
B_{i} =   I - a_i^2 A_{i} \, ,
\ee
\be
B_{ilm..} = A_{lm..} - a_i^2 A_{ilm..} \ .  
\ee

By using these relations we obtain the final forms of 
$p_i$, $q_i$ and ${\cal I}_{10}$, which only contain the quantities  
\be 
A_i , \overline A_{12}, \overline A_{122}, \overline A_{112}\, . 
\ee
Moreover, the $A_i$'s are not independent, 
since $A_3 = 2 - A_1 - A_2$. 
The derivatives of these quantities have been evaluated 
analytically and are reported in the next section. 
In axisymmetry, we have $A_3 = 2 - 2 A_1$ and 
\bea 
\label{oveas1}
\overline A_{12} & = & { 1 \over 2} + { 1 \over 4 
e^2} \left ( 3A_1 - 2 \right )  \, , \\ 
\label{oveas2}
\overline A_{112} &=&
\overline A_{122} =
{ 5\over 6 } \left [ { 1 \over 2} \left ( 1 +
{ 1 - e^2  \over 5 e^2 } \right )
+ { 1 \over 4 e^4 }
\left (
3 A_1 - 2 \right ) \right ]\, . 
\eea

\subsection{{\it Some Useful Derivatives}}

The first 
and second derivatives of the quantities $A_i$, $\overline A_{12}$, 
$\overline A_{122}$, $\overline A_{112}$ with respect to the axial ratios 
have been evaluated using the chain rule 
\be 
\der{A_{ijl}}{\lambda_i} = \sum_k \der{A_{ijl..}}{a_k} \der {a_k}
{\lambda_i} \, , 
\ee
and exploiting the definitions (\ref{defis}). 
The resulting expressions have been evaluated in the axisymmetric case, 
and substituted into the variations of the conserved quantities by using MAPLE. 
The final results, already specialized to axisymmetry, are 
\bea
\nonumber
\der{A_1}{\lambda_1}& = &
{\frac {15\,A_1+12\,A_1
{{  \lambda}}^{3}-18\,{{  \lambda}}^{3}}{\left (8-8\,{{  \lambda}}^{3}
\right ){  \lambda}}}
\\
\nonumber
\der{A_1}{\lambda_2}& = &
{\frac {-3\,A_1+12\,A_1
{{  \lambda}}^{3}-6\,{{  \lambda}}^{3}}{\left (8-8\,{{  \lambda}}^{3}\right 
){  \lambda}}}
\\
\nonumber
\der{A_2}{\lambda_1}& = &
{\frac {-3\,A_1+12\,A_1{{  \lambda}}
^{3}-6\,{{  \lambda}}^{3}}{\left (8-8\,{{  \lambda}}^{3}\right 
){  \lambda}}}
\\
\nonumber
\der{A_2}{\lambda_2}& = &
{\frac {15\,A_1+12\,A_1{{  \lambda}}^{3}-18\,{{  \lambda}}^{3}}
{\left (8-8\,{{  \lambda}}^{3}
\right ){  \lambda}}}
\\
\nonumber
\der{\overline A_{12}} {\lambda_1}& = & -{\frac {9\,A_1
+18\,{{  \lambda}}^{3}+12\,{{  
\lambda}}^{6}-54\,A_1{{  \lambda}}^{3}}{16\,\left (1-
{{  \lambda}}^{3}\right )^{2}{  \lambda}}}
\\ \label{dpri} & & 
\\ 
\nonumber
\der{\overline A_{12}} {\lambda_2}& = & 
{\frac {27\,A_1-42\,{{  \lambda}}^{3}+12\,{{  
\lambda}}^{6}+18\,A_1{{  \lambda}}^{3}}{16\,\left (1-
{{  \lambda}}^{3}\right )^{2}{  \lambda}}}
\\ 
\nonumber
\der{\overline A_{122}}{\lambda_1}& = &
{\frac {-615\,A_1+306\,{{  \lambda}}^{3}-1336\,{
{  \lambda}}^{6}+400\,{{  \lambda}}^{9}+1560\,A_1{{
  \lambda}}^{3}}{384\,\left (1-{{  \lambda}}^{3}\right )^{3}{  \lambda}}}
\\ 
\nonumber
\der{\overline A_{122}}{\lambda_2}& = &
-{\frac {-1095\,A_1+2034\,{{  \lambda}}^{3}-1448\,
{{  \lambda}}^{6}+464\,{{  \lambda}}^{9}-480\,A_1{{
  \lambda}}^{3}}{384\,\left (1-{{  \lambda}}^{3}\right )^{3}{  \lambda}}}
\\ 
\nonumber
\der{\overline A_{112}}{\lambda_1}& = &
{\frac {15\,A_1-738\,{{  \lambda}}^{3}-424\,{{
  \lambda}}^{6}+112\,{{  \lambda}}^{9}+1560\,A_1{{
  \lambda}}^{3}}{384\,\left (1-{{  \lambda}}^{3}\right )^{3}{  \lambda}}}
\\ 
\nonumber
\der{\overline A_{112}}{\lambda_2}& = &
-{\frac {-465\,A_1+990\,{{  \lambda}}^{3}-536\,{{
  \lambda}}^{6}+176\,{{  \lambda}}^{9}-480\,A_1{{
  \lambda}}^{3}}{384\,\left (1-{{  \lambda}}^{3}\right )^{3}{  
\lambda}}}\, , 
\eea
and 
\bea
\nonumber
\derss{A_1}{\lambda_1}& = &
{\frac {39\,A_1-306\,{{  \lambda}}^{3}+516\,A_1{{  \lambda}}
^{3}+120\,A_1{{  \lambda}}^{6}-144\,{{  \lambda}}^{6}}{32\,\left (1-{{  \lambda}}^{3}\right 
)^{2}{{  \lambda}}^{2}}}
\\
\nonumber
\ders{A_1}{\lambda_1}{\lambda_2} & = &
{\frac {-9\,A_1+72\,A_1
{{  \lambda}}^{3}-18\,{{  \lambda}}^{3}-72\,{{  \lambda}}^{6}+72\,{  A1}({
  \lambda},{  l2}){{  \lambda}}^{6}}{32\,\left (1-{{  \lambda}}^{3}\right )^
{2}{{  \lambda}}^{2}}}
\\
\nonumber
\derss{A_2}{\lambda_1}& = &
{\frac {3\,A_1+6\,{{  \lambda}}^{3}-96\,{{  \lambda}
}^{6}+12\,A_1{{  \lambda}}^{3}+120\,A_1
{{  \lambda}}^{6}}{32\,\left (1-{{  \lambda}}^{3}\right )^
{2}{{  \lambda}}^{2}}}
\\
\nonumber
\ders{A_2}{\lambda_1} {\lambda_2}& = &
{\frac {-9\,A_1+72\,A_1{{  \lambda}}^{3}
-18\,{{  \lambda}}^{3}-72\,{{  \lambda}}^{6}+72\,{  A1}({
  \lambda},{  l2}){{  \lambda}}^{6}}{32\,\left (1-{{  \lambda}}^{3}\right )^
{2}{{  \lambda}}^{2}}}
\\ 
\nonumber
\derss {\overline A_{12}} {\lambda_1} & =& -{\frac {-117 A_1 +144 A_1
{{  \lambda}}^{3}-234{{  \lambda}}^{3}
+3048 {{  \lambda}}^{6}+336 {{
  \lambda}}^{9}-4752
A_1{{  \lambda}}^{6}}{256\,
\left (1-{{  \lambda}}^{3}\right )^{3}{{  \lambda}}^{2}}}
\\ 
\label{dsec} & & 
\\ 
\nonumber
\ders {\overline A_{12}} {\lambda_1}{\lambda_2} & = &
-{\frac {-195\,A_1-960\,A_1
{{  \lambda}}^{3}+634\,{{  \lambda}}^{3}+1144\,{{  \lambda}}^{6}+112\,{{
  \lambda}}^{9}-1680\,A_1{{  \lambda}}^{6}}{256\,
\left (1-{{  \lambda}}^{3}\right )^{3}{{  \lambda}}^{2}}}
\\ 
\nonumber
\derss{\overline A_{122}}{\lambda_1}& = &
{\frac {8325\,A_1-20220\,A_1
{{  \lambda}}^{3}-7926\,{{  \lambda}}^{3}+18624\,{{  \lambda}}^{6}-
21296\,{{  \lambda}}^{9}+4928\,{{  \lambda}}^{12}+20400\,A_1
{{  \lambda}}^{6}}{512\,{{  \lambda}}^{2}\left (1-{{  \lambda}}^{3}
\right )^{4}}}
\\ 
\nonumber
\ders{\overline A_{122}}{\lambda_1}{\lambda_2}& = &
-{\frac {5535\,A_1-10440\,A_1
{{  \lambda}}^{3}-7362\,{{  \lambda}}^{3}+21528\,{{  \lambda}}^{6}-
11184\,{{  \lambda}}^{9}+2688\,{{  \lambda}}^{12}-3600\,A_1
{{  \lambda}}^{6}}{512\,{{  \lambda}}^{2}\left (1-{{  \lambda}}^{3}
\right )^{4}}}
\\ 
\nonumber
\derss{\overline A_{112}}{\lambda_1}& = &
{\frac {-75\,A_1-480\,A_1
{{  \lambda}}^{3}+1386\,{{  \lambda}}^{3}-12424\,{{  \lambda}}^{6}-2704\,{{
  \lambda}}^{9}+20400\,A_1{{  \lambda}}^{6}+512\,{{
  \lambda}}^{12}}{512\,{{  \lambda}}^{2}\left (1-{{  \lambda}}^{3}\right )^{4}
}}
\\ 
\nonumber
\ders{\overline A_{112}}{\lambda_1}{\lambda_2}& = &
-{\frac {-345\,A_1-4560\,A_1
{{  \lambda}}^{3}+2382\,{{  \lambda}}^{3}+3272\,{{  \lambda}}^{6}+16\,{
{  \lambda}}^{9}-3600\,A_1{{  \lambda}}^{6}}{512\,{{
  \lambda}}^{2}\left (1-{{  \lambda}}^{3}\right )^{4}}}
\eea

\clearpage

\figcaption[fig1.ps]{
The ratio $\Omega^2 / \left ( \pi \rho_0 \right ) $ versus the 
eccentricity for PN equilibrium sequences of constant rest mass (solid 
lines). 
The different curves correspond to 11 equally spaced values of the 
compaction parameter $M/R_s$ in the range [0., 0.275]. This parameter 
characterizes the nonrotating spherical member of each sequence. The dashed 
line is the Newtonian Maclaurin sequence ($M/R_s=0$).\label{fig1}}

\figcaption[fig2.ps]{The ratio $T/|W|$ versus the eccentricity for the PN 
equilibrium sequence with $ \left (M/R_s \right)_{max}= 5/18$ (solid 
line), and the Newtonian Maclaurin sequence (dashed line).\label{fig2}}

\figcaption[fig3.ps]{
The ratio $\Omega^2 / \left ( \pi \rho_0 \right ) $ versus $T/|W|$ for 
the same sequences as in Figure 1 (solid lines); the dashed line is
the Newtonian Maclaurin sequence. Squares mark the secular instability point.
\label{fig3}.}

\figcaption[fig4.ps]{Comparison between the ellipsoidal PN function $E(e)$ 
(solid line) and the 
corresponding function $E_{Ch}$ derived 
by Chandrasekhar (1965b, dotted line)\label{fig4}.}

\figcaption[fig5.ps]{The ellipsoidal PN equilibrium sequence (solid line) 
and the Chandrasekhar (1965b) 
sequence (dotted line) for $2 M_c /a_1 = 0.206$. Squares represent the 
Butterworth \& 
Ipser (1976) numerical values, 
and the dashed line is the Newtonian Maclaurin sequence.\label{fig5}}

\figcaption[fig6.ps]{The critical ratio $T/ |W|$ at the secular 
instability point versus the compaction parameter $M / R_s$ along a PN 
equilibrium sequence (solid line). The dashed line is the Newtonian 
value.\label{fig6}}

\clearpage

%

\begin{deluxetable}{ccccccccc}
\tablecolumns{9}
\tablewidth{0pc}
\tablecaption{The Function $E(e)$; Comparison with 
Chandrasekhar (1965b) \label{tab1}}
\tablenum{1}
\tablehead{
\colhead{ $e$ } & 
\colhead{$E \left ( e \right )$}&
\colhead{$E_{Ch} \left( e \right )$} & 
\colhead{err\tablenotemark{a}} & 
\vline & 
\colhead{ $e$ } & 
\colhead{$E \left ( e \right )$}&
\colhead{$E_{Ch} \left( e \right )$} & 
\colhead{err\tablenotemark{a}} \nl} 
\startdata
                  0    & 0      & 0       & -- & \vline & 
                  0.65 & 0.1191 & 0.1202 & 0.0094 \\                 
                  0.20 & 0.0090 & 0.0093 & 0.0397 & \vline & 
                  0.70 & 0.1448 & 0.1444 & 0.0026 \\ 
                  0.25 & 0.0142 & 0.0147 & 0.0389 & \vline & 
                  0.75 & 0.1752 & 0.1719 & 0.0195 \\ 
                  0.30 & 0.0207 & 0.0215 & 0.0379 & \vline & 
                  0.80 & 0.2118 & 0.2026 & 0.0447 \\ 
                  0.35 & 0.0287 & 0.0298 & 0.0366 & \vline & 
       0.8127\tablenotemark{\ast} & 0.2223 & 0.2108 & 0.0531 \\ 
                  0.40 & 0.0383 & 0.0397 & 0.0350 & \vline & 
                  0.85 & 0.2564 & 0.2355 & 0.0847 \\ 
                  0.45 & 0.0480 & 0.0513 & 0.0661 & \vline & 
                  0.90 & 0.3115 & 0.2666 & 0.1552 \\ 
                  0.50 & 0.0629 & 0.0650 & 0.0268 & \vline &  
  0.9529\tablenotemark{\ast \ast } & 0.3838 & 0.2768 & 0.3240 \\
                  0.55 & 0.0790 & 0.0808 & 0.0220 & \vline &
                  0.96 & 0.3939 & 0.2723 & 0.3651 \\ 
                  0.60 & 0.0974 & 0.0991 & 0.0177 & \vline & & & & \\
\enddata
\tablenotetext{a}{The fractional difference is defined as 
$$
{\rm err}  \equiv  2 { \left | E \left ( e \right ) - 
E_{Ch} \left( e 
\right ) \right | \over  E \left ( e \right ) + E_{Ch} \left( e \right  ) }
\, . $$} 
\tablenotetext{\ast}{Newtonian value for the secular instability point.} 
\tablenotetext{\ast \ast }{Newtonian value for the dynamical instability 
point.} 
\end{deluxetable}

\clearpage

%

\begin{deluxetable}{ccccccc}
\tablecolumns{7}
\tablewidth{0pc}
\tablecaption{Comparison with Butterworth \& Ipser (1976); 
$2 M_c /a_1=0.206$\tablenotemark{a} {\label{tab2}}}
\tablenum{2}
\tablehead{
\colhead{ $e$ } & 
\colhead{$e_{BI}$} &
\colhead{${ \Omega^2 \over \pi \rho_0}$} & 
\colhead{$\left ( { \Omega^2 \over \pi \rho_0} \right)_{BI}$} & 
\colhead{${ \Omega^2 \over \pi \rho_0} - \left({ \Omega^2 \over \pi \rho_0}
\right)_N $} & 
\colhead{$\left [ { \Omega^2 \over \pi \rho_0} - \left({ \Omega^2 \over \pi 
\rho_0} \right)_N \right ] _{BI}$ }& 
\colhead{err \tablenotemark{b}} \nl }
\startdata
         0.199 & 0.200  & 0.024  & 0.025 & 
         0.002 & --  & -- \\                 
         0.298 & 0.300  & 0.054  & 0.058 & 
         0.006 & --  & -- \\                 
         0.397 & 0.400  & 0.097  & 0.107 & 
         0.011 & --  & -- \\                 
         0.498 & 0.500  & 0.155  & 0.164 & 
         0.018 & --  & -- \\                 
         0.598 & 0.600  & 0.227  & 0.247 & 
         0.027 & --  & --\\                 
         0.695 & 0.700  & 0.314  & 0.329 & 
         0.040 & 0.051 & 0.24 \\                 
         0.796 & 0.800  & 0.419  & 0.429 & 
         0.059 & 0.066 & 0.10 \\                 
         0.846 & 0.850  & 0.472  & 0.474 & 
         0.072 & 0.066 & 0.08 \\                 
\enddata
\tablenotetext{a}{Corresponding to $\gamma_s = 0.154$, where $\gamma_s$ is the 
parameter used by Butterworth \& Ipser (1976).} 
\tablenotetext{b}{The fractional difference with respect the Butterworth \& Ipser value is 
defined as 
$$
{\rm err}  \equiv  2 { \left | Q - Q_{BI} \right | 
\over  Q + Q_{BI}}
\, , $$
where $Q = \Omega^2 /\left ( \pi
\rho_0 \right ) - \left [ \Omega^2 /\left ( \pi \rho_0 \right ) \right ]_N$.}
\end{deluxetable}

\clearpage
%

\begin{deluxetable}{cccc}
\tablecolumns{4}
\tablewidth{0pc}
\tablecaption{Location of the PN secular 
instability point{\label{tab3}}} 
\tablenum{3}
\tablehead{
\colhead{ ${ M \over R_s}$ } & 
\colhead{$e_{sec}$} &
\colhead{$\left ({ \Omega^2 \over \pi \rho_0} \right)_{sec}$} & 
\colhead{$ \left ( { T \over |W| } \right )_{sec}$ } \nl }
\startdata
         0.\tablenotemark{a} & 0.8127  & 0.3742  & 0.1375 \\ 
         0.025 & 0.8561  & 0.4230  & 0.1673 \\ 
         0.050 & 0.8850  & 0.4582  & 0.1924 \\ 
         0.075 & 0.9042  & 0.4831  & 0.2120 \\ 
         0.100 & 0.9171  & 0.5011  & 0.2268 \\ 
         0.125 & 0.9261  & 0.5144  & 0.2378 \\ 
         0.150 & 0.9325  & 0.5247  & 0.2458 \\ 
         0.175 & 0.9372  & 0.5327  & 0.2516 \\ 
         0.200 & 0.9403  & 0.5384  & 0.2554 \\ 
         0.225 & 0.9425  & 0.5424  & 0.2578 \\ 
         0.250 & 0.9436  & 0.5446  & 0.2591 \\ 
         0.275 & 0.9441  & 0.5457  & 0.2597 \\ 
\enddata
\tablenotetext{a}{Newtonian limit.}
\end{deluxetable}

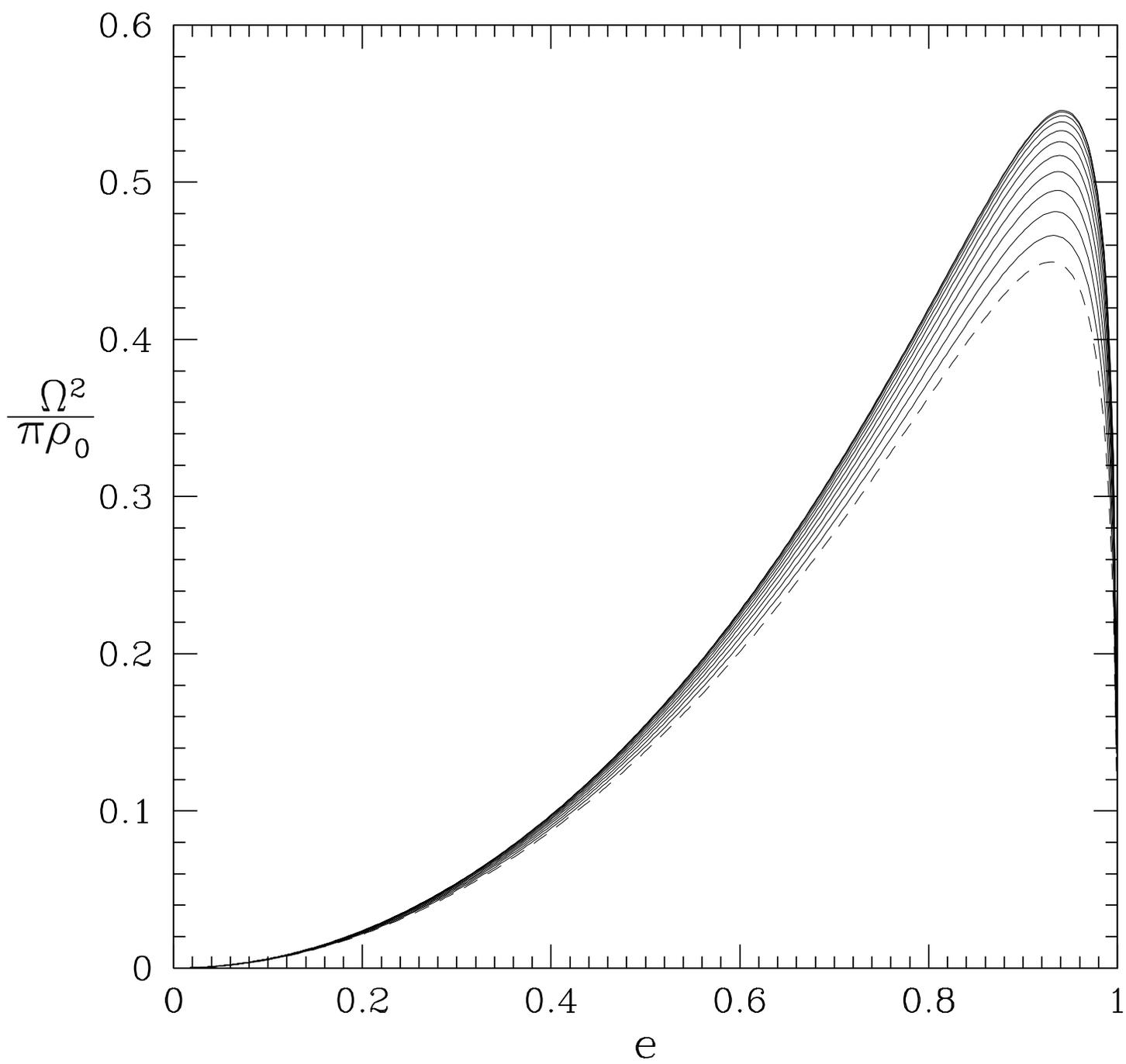
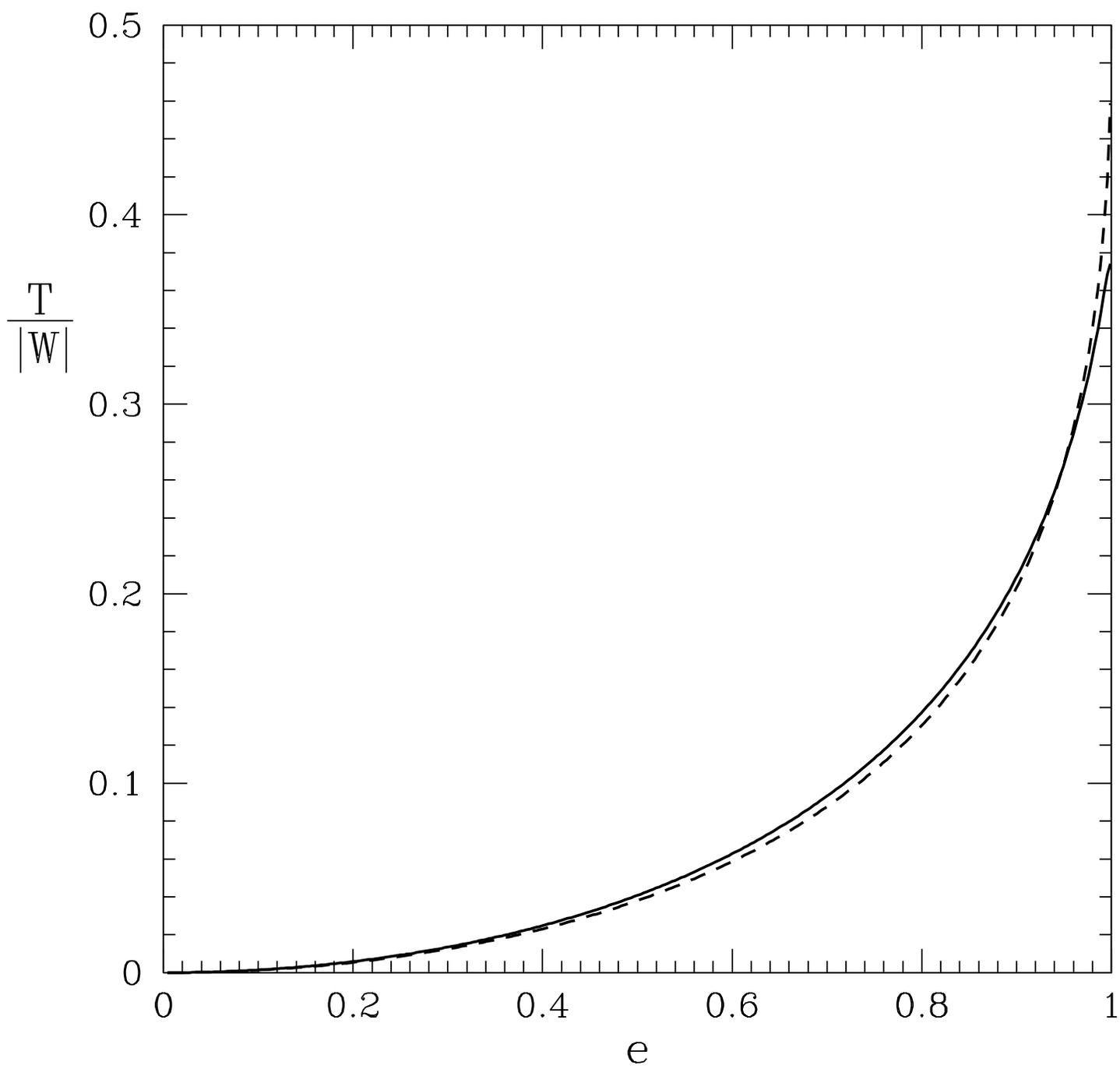
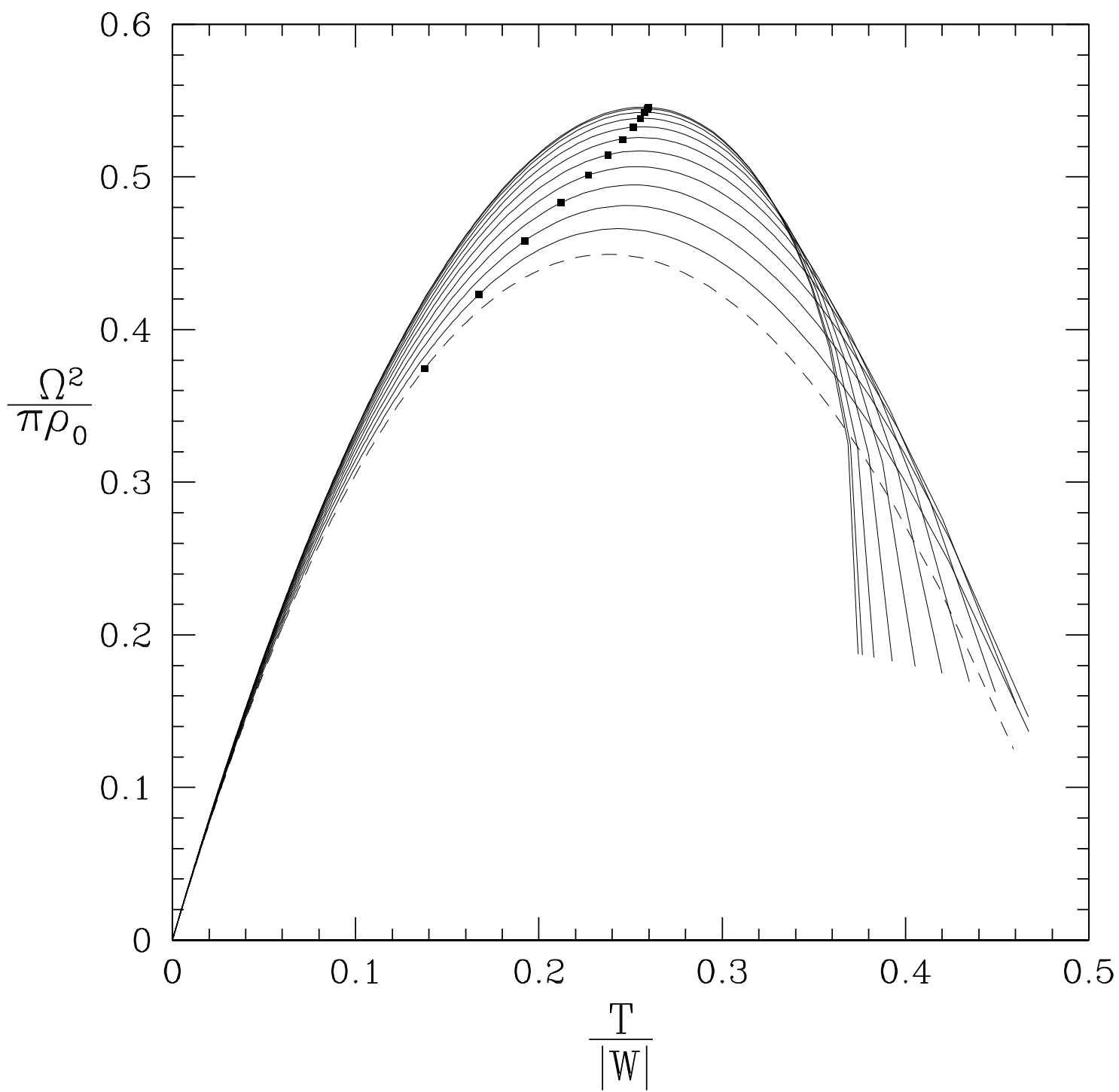
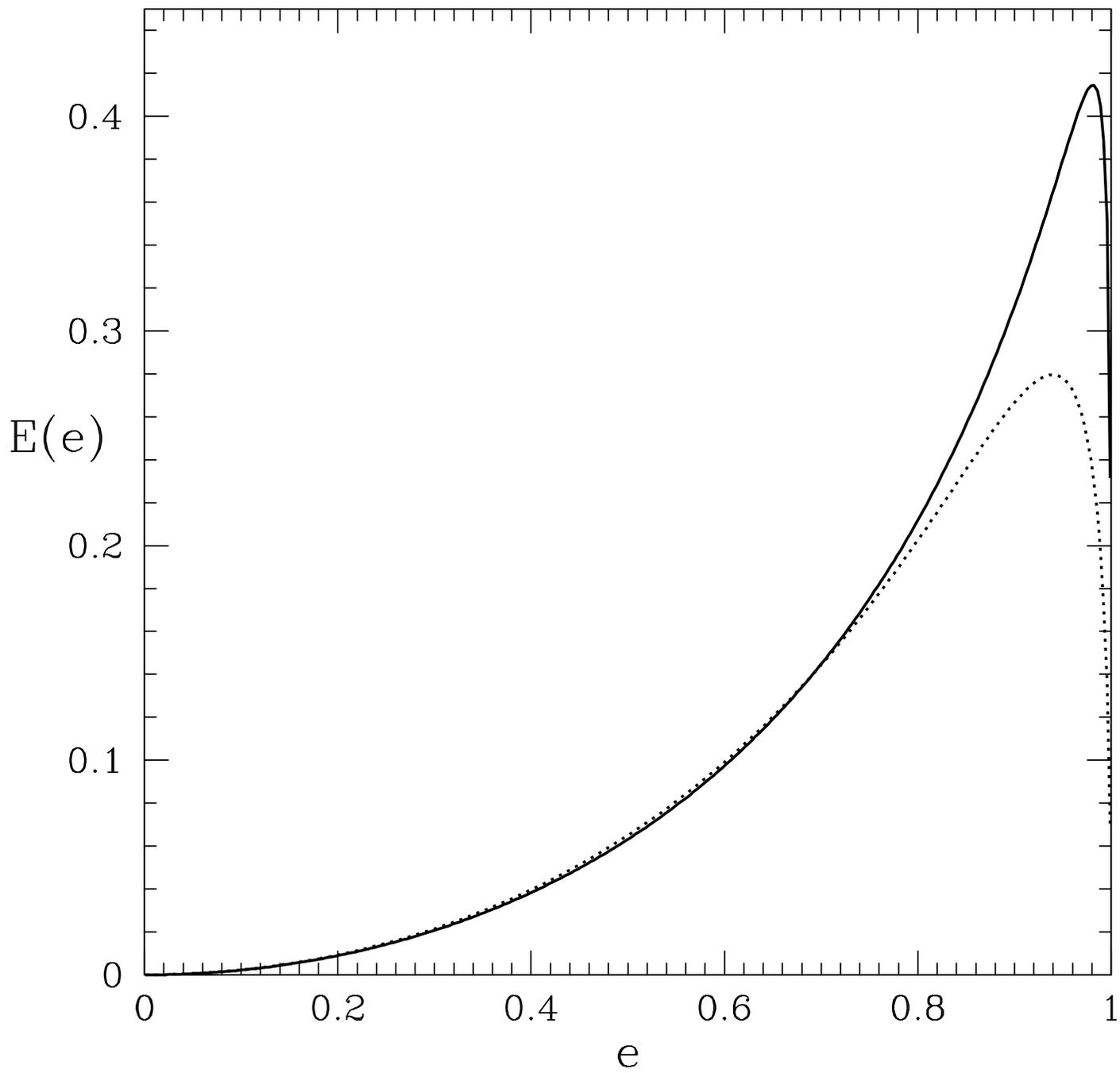
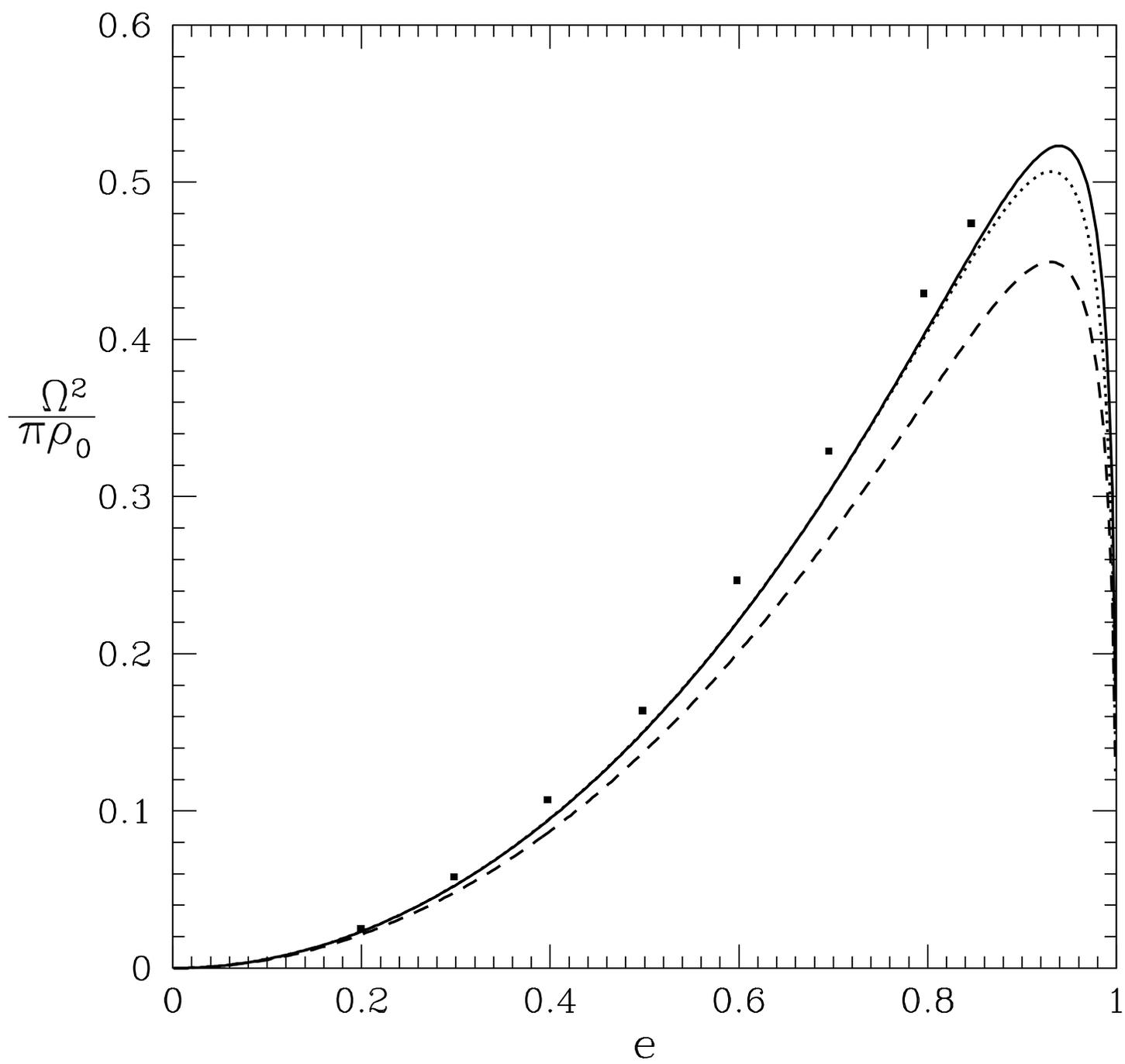
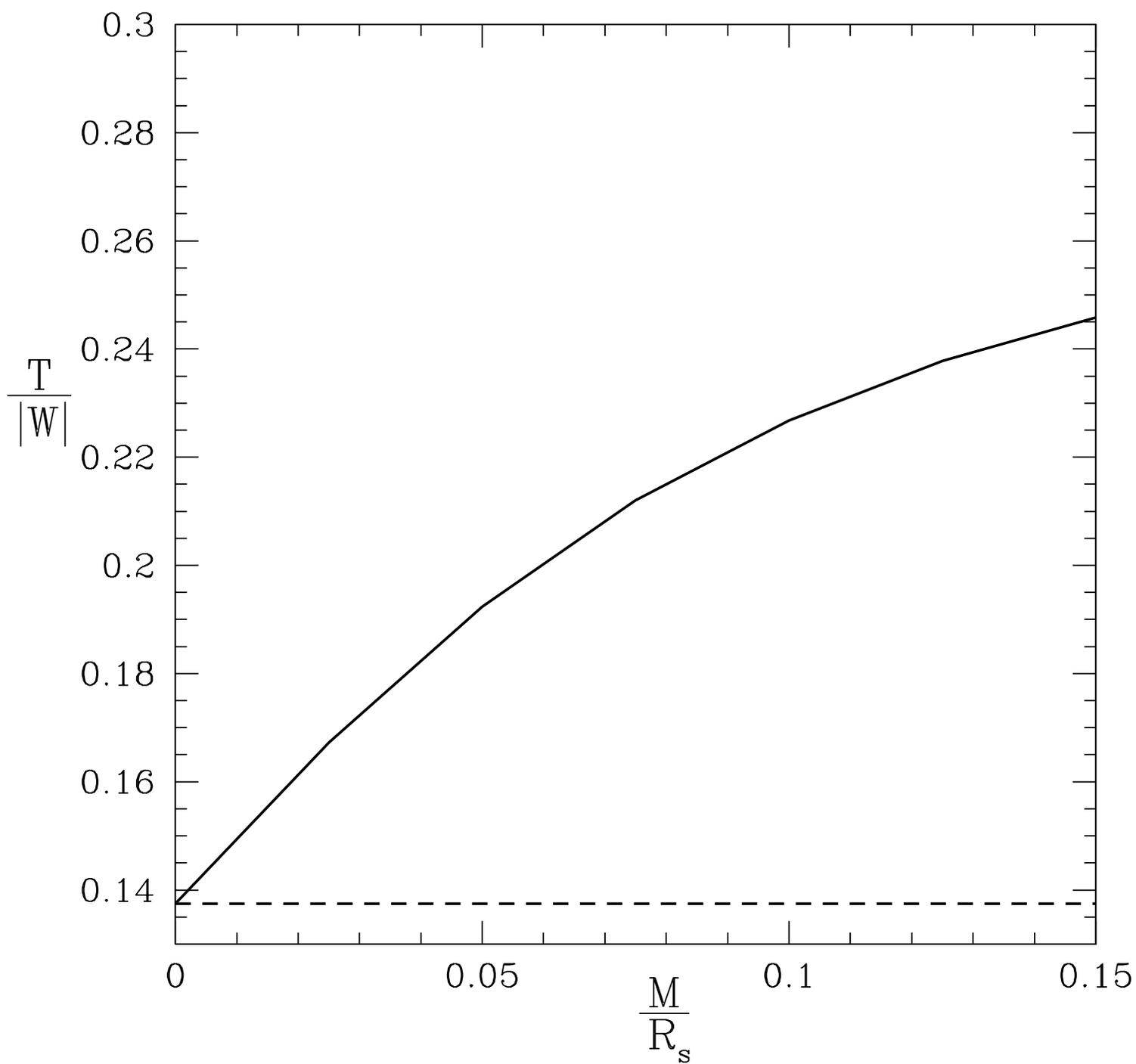
\end{document}